\documentclass[11pt]{article}
\pdfoutput=1
\usepackage{amsfonts, amsthm, simplewick, datetime, multirow, array, booktabs, verbatim, floatrow, caption, subcaption,url, latexsym, graphicx, graphics, amsmath, amssymb, slashed,mathabx, color}
\usepackage[font=footnotesize, labelfont={sf,bf}, margin=1cm]{caption}
\usepackage{jcappub,  hyperref} 

\newcommand{\nc}{\newcommand}

\nc{\beq}{\begin{equation}}
\nc{\eeq}{\end{equation}}
\nc{\beqa}{\begin{eqnarray}}
\nc{\eeqa}{\end{eqnarray}}
\nc{\bea}{\begin{eqnarray}}
\nc{\eea}{\end{eqnarray}}
\nc{\ra}{\rightarrow}
\nc{\Tr}{{\rm Tr}}
\nc{\slsh}{\slash\hspace*{-0.22cm}}

\def\be{\begin{equation}}
\def\ee{\end{equation}}
\def\bea{\begin{eqnarray}}
\def\eea{\end{eqnarray}}
\def\bit{\begin{itemize}}
\def\eit{\end{itemize}}

\nc{\barray}{\begin{eqnarray}}
\nc{\earray}{\end{eqnarray}}
\nc{\barrayn}{\begin{eqnarray*}}
\nc{\earrayn}{\end{eqnarray*}}
\nc{\mc}{\mathcal}
\nc{\M}{\mathcal{M}}

\nc{\h}{$h$}
\nc{\infinity}{\infty}

\def\ben{\begin{enumerate}}
\def\een{\end{enumerate}}

\newcommand{\cref}[1]{Chapter~\ref{c.#1}}

\newcommand{\neff}{N_{\mathrm{eff}}}

\def\to{\rightarrow}

\renewcommand\thesection{\arabic{section}}

\makeatletter
\def\p@subsection{}
\def\p@subsubsection{}
\makeatother

\addtocounter{page}{2}

\begin{document}

\title{
Singularities in the Gravitational Capture of Dark Matter through Long-Range Interactions
}

\author{Cristian Gaidau}
\author{and Jessie Shelton}

\affiliation{Illinois Center for the Advanced Study of the Universe, \\Department of Physics, University of Illinois at Urbana-Champaign,\\  1110 W Green St., Urbana, IL 61801, USA}

\emailAdd{gaidau2@illinois.edu}
\emailAdd{sheltonj@illinois.edu}

\date{\today}

\abstract{
We re-examine the gravitational capture of dark matter (DM) through long-range interactions. We demonstrate that neglecting the thermal motion of target particles, which is often a good approximation for short-range capture, results in parametrically inaccurate results for long-range capture. When the particle mediating the scattering process has a mass that is small in comparison to the momentum transfer in scattering events, correctly incorporating the thermal motion of target particles results in a quadratic, rather than logarithmic, sensitivity to the mediator mass, which substantially enhances the capture rate.  We quantitatively assess the impact of this finite temperature effect on the captured DM population in the Sun as a function of mediator mass. We find that capture of DM through light dark photons, as in e.g. mirror DM, can be
powerfully enhanced, with self-capture attaining a geometric limit over much of parameter space.  
For visibly-decaying dark photons, thermal corrections are not large in the Sun, but may be important in understanding long-range DM capture in more massive bodies such as Population III stars. We additionally provide the first calculation of the long-range DM self-evaporation rate.
}

\maketitle

 \setcounter{equation}{0} \setcounter{footnote}{0}


\section{Introduction}
\label{s.intro}

The gravitational capture of dark matter (DM) in massive bodies such as stars and planets is a powerful probe of DM's non-gravitational interactions with the SM.  Dark matter that undergoes scattering with the material in a massive body can lose enough energy in the scattering event to become gravitationally bound.  Through such processes the massive body can build up a population of captured dark matter, which can in turn give rise to a variety of observational signatures depending on the detailed properties of DM.  If the captured DM can self-annihilate, then its annihilation products may be detected either directly, if they escape the body \cite{Silk:1985ax,Schuster:2009fc,Batell:2009zp}, or indirectly, through the energy they deposit into the capturing body \cite{Krauss:1985aaa,Kouvaris:2007ay,Mack:2007xj,Bertone:2007ae,Baryakhtar:2017dbj,Leane:2020wob}. Even in the absence of annihilations, the captured DM population may still be detectable through its gravitational effect on the capturing body \cite{Spergel:1984re,Goldman:1989nd, Lopes:2002gp,deLavallaz:2010wp,Frandsen:2010yj,Cumberbatch:2010hh,Taoso:2010tg,McDermott:2011jp}. 

 In computing the DM capture rate for non-degenerate systems, e.g., main sequence stars such as our Sun, planets such as our Earth, and nuclei in white dwarfs, it is common to neglect the temperature of the target particles. 
For short-range interactions, incorporating the finite temperature of the targets gives corrections of less than 10\% to the zero-temperature nuclear capture rate in the Sun for DM heavier than the evaporation mass \cite{Gould:1987ir, Busoni:2013kaa}, and thus in this case the zero-temperature approximation for DM capture is usually well justified.  However, we show here that for long-range interactions, the thermal motion of target nuclei has a far more dramatic effect on the capture rate.  The zero-temperature capture rate through long-range interactions is logarithmically divergent \cite{Fan:2013bea}, and is thus only logarithmically sensitive to the (model-dependent) regulator, which could be a screening scale \cite{Fan:2013bea} or an intrinsic mediator mass  \cite{Dasgupta:2020dik}. 
We find that retaining the finite temperature of the target nuclei instead results in a {\em quadratically} divergent capture rate in the long-range regime, resulting in both a substantially enhanced capture rate and increased sensitivity to the physics responsible for regulating the capture rate in the infrared.   Similar results hold for the self-capture of DM through long-range interactions.  

Capture via long-range interactions is particularly relevant for DM that interacts with itself and/or the SM via a new dark force.  DM self-interactions in such models can lead to novel cosmological and/or astrophysical signatures \cite{Feng:2009mn,Feng:2009hw,Loeb:2010gj,Cyr-Racine:2012tfp,Fan:2013yva,Curtin:2019ngc}.  These models generally feature secluded DM annihilations into mediators that may be long-lived, thus providing novel cosmic ray signatures \cite{Schuster:2009fc,Batell:2009zp,Schuster:2009au,Meade:2009mu,Bell:2011sn,Ajello:2011dq,Feng:2015hja,Feng:2016ijc,Adrian-Martinez:2016ujo,Leane:2017vag,Arina:2017sng,Ardid:2017lry,Niblaeus:2019gjk,Bell:2021pyy}.  Thermal target motion is known to be important in cases where short-range scattering on electrons dominates gravitational capture in the Sun, due to the much larger thermal velocities of electrons compared to nuclei \cite{Garani:2017jcj}.  The effect we point out in this paper is qualitatively distinct, and applies for long-range interactions between DM and any non-degenerate population of targets, including electrons.   For concreteness, we illustrate the general point with numerical results for the familiar case of nuclear capture in the Sun, which offers the most immediate detection prospects, but discuss how our results apply to other systems.

In Secs.~\ref{s.nuclear_cap} and~\ref{s.selfcaptAndselfejec} we discuss long-range nuclear capture and long-range DM self-capture, respectively. Both sections start with a review of the zero-temperature calculation, followed by the finite-temperature calculation incorporating the thermal motion of target particles. We quantify the impact of the finite target temperature and discuss its observational and phenomenological implications in Sec.~\ref{s.DiscussionImplications}. We conclude in Sec.~\ref{s.concl}, and provide additional technical details for the computation of nuclear and self-capture rates in  Appendices~\ref{appendix.A} and~\ref{appendix.B}, respectively.   Finally, we discuss DM self-evaporation via long-range forces in Appendix~\ref{appendix.C}.


\section{Nuclear capture with thermal targets}
\label{s.nuclear_cap}

We begin with the calculation of the rate of DM gravitational capture in the Sun by scattering against nuclei, given by the nuclear capture coefficient $C_\mathrm{c}$. This rate depends on three ingredients: a model of the DM halo in the vicinity of the Sun, a model of the solar interior, and a model for the DM-nuclear interaction cross-section.

\par
For DM particles in the halo we assume a Maxwell distribution of velocities,
\begin{align}
f(u)d^3u = 4\pi\left(\frac{3}{2\pi\bar{v}^2}\right)^{3/2}\exp{\left(-\frac{3u^2}{2\bar{v}^2}\right)}u^2du,
\end{align}
where $u$ is the DM speed far away from the Sun and $\bar{v}^2$ is the DM rms speed. To find the distribution of initial DM speeds as seen by the Sun, we perform a Galilean transformation to the Sun's rest frame and average over the solid angle:
\begin{align}
f_{\eta}(u) = \left(\frac{3}{2\pi \bar{v}^2}\right)^{3/2}\exp{\left(-\frac{3(u^2+\tilde{v}^2)}{2\bar{v}^2}\right)}\frac{\sinh{(\frac{3}{2\bar{v}^2}2u\tilde{v})}}{\frac{3}{2\bar{v}^2}2u\tilde{v}},
\label{fEta}
\end{align}
where $\tilde{v}$ is the Galilean boost from the Galactic rest frame to the Sun rest frame.  Assuming a virialized DM halo, $\bar{v}^2 = 3/2 \,v^2_{\text{LSR}}$, where $v_{\text{LSR}}$ is the Local Standard of Rest at the position of the Sun. Refs.~\cite{bovy2009galactic, reid2009trigonometric, mcmillan2010uncertainty} estimate $v_{\text{LSR}} = 235 \text{ km}/\text{s}$, so that $\bar{v} = 288 \text{ km}/\text{s}$. Given the Sun's peculiar velocity ${\bf{v}}_{\text{pec}}=(11,12,7) \text{ km}/\text{s}$~\cite{schonrich2010local}, it moves at $\tilde{v} = 247 \text{ km}/\text{s}$ with respect to the DM halo.  Meanwhile, the number density of dark matter in the solar neighborhood is $n_{\text{DM}} = \rho_{\text{DM}}/M_{\text{DM}}$, where we take $\rho_{\text{DM}}=0.4\,\mathrm{GeV}/\mathrm{cm}^3$~\cite{Read:2014qva}. 

We model the density and composition of the Sun using the BS05(AGS, OP) solar model~\cite{Bahcall:2004pz}.  For DM-nuclear interactions, we consider for concreteness a minimal reference model that consists of a fermionic DM particle $\chi$ with mass $M_{\text{DM}}$, which couples to a massive $U(1)$ gauge boson $A_D$ with mass $M_{\gamma_D}$. We take $A_D$ to couple to the SM via kinetic mixing with the hypercharge gauge boson~\cite{Holdom:1985ag}. The diagonalization of the gauge field kinetic terms results in the mostly-$A_D$ mass eigenstate picking up interactions with the SM via small mixings with hypercharge.  In this paper we will focus on a light $A_D$ boson, such that its couplings to the SM are photon-like. The Lagrangian in the mass basis then reads
\begin{align}
\mathcal{L} = -\frac{1}{4}F_{D\mu\nu}F^{\mu\nu}_{D} +\frac{1}{2}M^2_{\gamma_D}A^2_{D} + \bar{\chi}(i\slashed{D}-M_{\text{DM}})\chi -\sum_f q_fe(A_{\mu} +\epsilon A_{D\mu})\bar{f}\gamma^{\mu}f,
\end{align}
where $\slashed{D} = \slashed{\partial} +ig_D\slashed{A}_{D}$, $\epsilon$ controls the kinetic mixing and $f$ is a SM fermion with electric charge $q_f$. For previous studies of gravitational capture in this model, see Refs.~\cite{Feng:2015hja,Feng:2016ijc,Bell:2021pyy}.

\subsection{Zero-temperature calculation}

We begin with a review of the zero-temperature calculation. We construct the capture coefficient following the procedure outlined in Gould's work~\cite{Gould:1987ju,Gould:1987ir,gould1988direct}, where the differential capture rate is integrated over spherical shells of the capturing body, 
\begin{align}
C_{\text{c}} = \sum_i C_{\text{c}, i} = \sum_i \int_0^R dr 4\pi r^2\frac{dC_{\text{c}, i}}{dV},
\end{align}
where $R$ is the radius of the Sun, and the sum runs over the contributions of different nuclear species. The capture rate per unit volume is
\begin{align}
\frac{dC_{\text{c},i}}{dV} = n_i(r)n_{\text{DM}}\int d^3\vec{u}\frac{w}{u}f_\eta(u)|\vec{w}|\sigma_{c, i},
\label{eqn2p5}
\end{align}
where $n_i$ is the number density of nuclear species $i$, $n_{\text{DM}}$ is the local DM halo number density, $\sigma_{c, i}$ is the DM-nucleus  cross section for scatterings that lead to capture, and $w^2 = u^2+v^2_{\text{esc}}(r)$ is the DM speed before the collision, reflecting the increase in kinetic energy as the incoming DM falls into the Sun's potential well. 

We will consider the scattering process in the center of mass (CM) frame.
Let $\vec{v}'_1$ and $\vec{v}_{n}^{CM}$ be the CM velocities of the DM particle and nucleus before the collision. The Galilean transformation to the CM frame reads:
\begin{align}
\vec{w} = \vec{v}'_1 +\vec{v}_{CM}, \phantom{space} \vec{v}_n = 0 = \vec{v}^{CM}_n +\vec{v}_{CM},
\end{align}
such that $m_i\vec{v}^{CM}_n + M_{\text{DM}}\vec{v}'_1 =0$. We note that
\begin{align}
\vec{v}_{CM} = \frac{M_{\text{DM}}}{m_i}\vec{v}'_1 \quad \text{and} \quad \vec{w} = \left(1+\frac{M_{\text{DM}}}{m_i}\right) \vec{v}'_1 .
\end{align}
The vectors $\vec{w}, \vec{v}_{CM}$ and $\vec{v}'_1$ are all collinear, which greatly simplifies the angular structure of the integral in Eq.~\ref{eqn2p5}. We now switch the integration variable to $d^3\vec{w}$, giving
\begin{align}
\frac{dC_{\text{c},i}}{dV} = n_i(r)n_{\text{DM}}\int d^3\vec{w}\Theta (|\vec{w}|-v_{\text{esc}}(r))f_\eta(\sqrt{w^2-v^2_{\text{esc}}(r)})|\vec{w}|\sigma_{c,i}.
\label{eqn:Cc320}
\end{align}

\par
We turn to the construction of the capture cross-section $\sigma_{c, i}$. DM scattering off nuclei proceeds via the $t$-channel exchange of $A_{D}$. In the non-relativistic limit, the matrix element-squared for this process reads
\begin{align}
|\mathcal{M}|^2 = \frac{4g_D^2\epsilon^2e^2}{v_1^{'4}}\frac{m_n^2}{M_{\text{DM}}^2}\frac{1}{(1-\cos\theta +\mu)^2},
\label{Eqn:MSqCc} 
\end{align}
where $\theta$ is the scattering angle in the CM frame, $m_n$ is the mass of the nucleon and $\mu$ regulates the forward singularity. For a massive dark gauge boson of mass $M_{\gamma_D}$, 
\begin{align}
\mu \equiv\frac{M^2_{\gamma_D}}{2M^2_{\text{DM}}v_1^{'2}}.
\label{Eqn:mu_defn}
\end{align}
In the case of coherent scattering, the nuclear cross-section can be simply written in terms of the nucleon cross-section
\begin{align}
\sigma_{c,i} = \left(\frac{\bar{m}_i}{\bar{m}}\right)^2Z_i^2\sigma_c,
\label{Eqn:sigmaNuclear}
\end{align}
where $Z_i$ is the number of protons in nucleus $i$, $\bar{m} \equiv M_{\text{DM}}m_n/( M_{\text{DM}}+m_n)$ is the reduced DM-nucleon mass and $\bar{m}_i \equiv M_{\text{DM}}m_i/( M_{\text{DM}}+m_i)$ is the reduced DM-nucleus mass. Using the matrix element in Eq.~\ref{Eqn:MSqCc} above, we now calculate $\sigma_c$:
\begin{align}
\sigma_c = \frac{1}{16m_n^2M_{\text{DM}}^2}\frac{\bar{m}^2}{(2\pi)^2}\int d\Omega_{f,\chi}|\mathcal{M}|^2\Theta\left(v_{\text{esc}}(r) - |\vec{v}^{lab}_{\text{DM} , f}|\right),
\label{Eqn:sigma_c}
\end{align}
where  $d\Omega_{f,\chi} = \sin\theta d\theta d\phi_{\chi} $ is the solid angle specifying the direction of the CM-frame velocity of the DM particle after the collision.  Here the step function enforces the kinematic requirement that the outgoing DM particle is gravitationally bound. Note that $\vec{v}_{\text{DM}, f}^{lab} = \vec{v}'_{1, f} +\vec{v}_{CM}$ and $|\vec{v}'_{1, f}|= |\vec{v}_1'|$. The step function thus imposes an upper bound on the $\cos\theta$ integral:
\begin{align}
\cos\theta \leq \cos\theta_+ \equiv \frac{m_i}{2M_{\text{DM}}}\left(\frac{v^2_{\text{esc}}(r)}{v^{'2}_1}-1 -\left(\frac{M_{\text{DM}}}{m_i}\right)^2\right) \leq 1,
\end{align}
where the second inequality is a consequence of the fact that $w\geq v_{\text{esc}}(r)$. For sufficiently energetic incoming DM particles, $\cos\theta_+<-1$ and gravitational capture cannot occur for any scattering angle $\theta$. Therefore, requiring $\cos\theta_+ \geq -1$ imposes an upper bound on $w$ for events leading to capture:
\begin{align}
w\leq v_{\text{esc}}(r)\frac{m_i+M_{\text{DM}}}{|m_i- M_{\text{DM}}|}.
\end{align}
The capture cross-section can be evaluated in closed form,
\begin{align}
\sigma_{c,i} = 2\pi Z_i^2\frac{\alpha\alpha_D\epsilon^2}{\bar{m}_i^2w^4}\frac{1+\cos\theta_+}{(2+\mu)(1-\cos\theta_++\mu)},
\end{align}
where $\alpha_D=g_D^2/(4\pi)$ and $\alpha=e^2/(4\pi)$. The kinematic integral now reads:
\begin{align}
\frac{dC_{\text{c},i}}{dV} = n_i(r)n_{\text{DM}}4\pi\int_{v_{\text{esc}}(r)}^{v_{\text{esc}}(r)\frac{m_i+M_{\text{DM}}}{|m_i-M_{\text{DM}}|}}dw f_\eta(\sqrt{w^2-v^2_{\text{esc}}(r)})w^3\sigma_{c,i}.
\label{Eqn:CcTzeroFinal}
\end{align}
We note that $\sigma_{c,i}$ is finite in the limit $\mu\rightarrow 0$ except at $w =v_{\text{esc}}(r)$, where $\cos\theta_{+}=1$. Upon integrating over the incoming DM speed $w$ in Eq.~\ref{Eqn:CcTzeroFinal}, this singularity in  $\sigma_{c,i}$ results in $C_{\text{c}}$ having a logarithmic sensitivity on the regulator $\mu$~\cite{Fan:2013bea, Chen:2015uha}. 

\par
Numerical calculations of $C_{\text{c}}$ with $T=0$ are shown as dashed lines in Fig.~\ref{plot:Ccapture}, where we examine the dependence of the capture coefficient on $M_{\gamma_D}$ at fixed $M_{\text{DM}}$ and vice-versa. Here we consider fixed couplings  $\alpha_D=10^{-3}$ and $\epsilon=10^{-3}$. 
\par
The left panel of Fig.~\ref{plot:Ccapture} illustrates two distinct regimes for the capture rate dependence on the mediator mass $M_{\gamma_D}$. When the dark photon mass is sufficiently large compared to the typical momentum transfer in a capture event, i.e., $\mu\gg 1$, the interaction is effectively pointlike, 
which in this model yields a constant, velocity-independent cross section proportional to $M_{\gamma_D}^{-4}$. In the opposite limit, when $M_{\gamma_D}$ is much smaller than the typical momentum transfer and $\mu \ll 1$, the  zero-temperature capture rate depends logarithmically on $M_{\gamma_D}$.

\subsection{Nuclear Capture at Finite Temperature}

We extend our previous treatment of nuclear capture by ``restoring'' the thermal motion of nuclear targets, which can be described by a Maxwell distribution 
\begin{align}
f_i(\vec{v}_n) = \left(\frac{m_i}{2\pi T_{\text{Sun}}}\right)^{3/2}\exp{\left(-\frac{m_iv_n^2}{2T_{\text{Sun}}}\right)}
\end{align}
in the Sun's rest frame, where $m_i$ is the mass of the nucleus, $v_n$ is the velocity of the nucleus and $T_{\text{Sun}}$ is the local temperature in the Sun. We take $T_{\text{Sun}}$ to be the (constant) core temperature for simplicity.

\par
In addition to the integral over the distribution of incoming DM speeds, the capture rate per unit shell now includes a convolution over the thermal distribution of the target particle velocity,
\begin{align}
\frac{dC_{\text{c}, i}}{dV} = n_i(r)n_{\text{DM}}\int d^3\vec{u}\frac{w}{u}f_{\eta}(u) \int d^3\vec{v}_n f_i(\vec{v}_n) |\vec{w} - \vec{v}_n|\sigma_{c,i}.
\end{align}

The scattering process itself remains simplest to describe in the CM frame; however, the kinematic conditions for capture depend on the outgoing velocities in the lab frame.  This complicates the evaluation of the final state phase space integrals, but as we will see being careful about the capture conditions is important for  obtaining the correct parametric dependence of the capture rate.
Transforming the velocity integrals $\int d^3\vec{u}\int d^3\vec{v}_n$ to the CM frame, we have for the incoming DM
\begin{align}
\int d^3\vec{u} \frac{w}{u}f_{\eta}(u) = \int d^3\vec{w}\Theta (|\vec{w}|-v_{\text{esc}}(r))f_{\eta}(\sqrt{w^2-v^2_{\text{esc}}(r)}).
\label{fTransform1}
\end{align}
Let $\vec{v}'_1$ and $\vec{v}_{n}^{CM}$ be the CM velocities of the DM particle and nucleus before the collision. The Galilean transformation reads
\begin{align}
\vec{w} = \vec{v}'_1 + \vec{v}_{CM},\phantom{space}\vec{v}_n = \vec{v}_{n}^{CM} + \vec{v}_{CM},
\label{GalTransform}
\end{align}
where
\begin{align}
\vec{v}_{CM} = \frac{m_i}{m_i+ M_{\text{DM}}}\vec{v}_n + \frac{M_{\text{DM}}}{m_i+ M_{\text{DM}}}\vec{w}.
\label{vCMEqn}
\end{align} 
Next we switch variables to $(\vec{v}_{CM}, \vec{v}'_1)$, where the Jacobian for this transformation is \\ 
$J=\left(1+M_{\text{DM}}/m_i\right)^3$, giving
\begin{align}
\frac{dC_{\text{c}, i}}{dV} &= n_i(r)n_{\text{DM}}\int d^3\vec{v}_{CM}\int d^3\vec{v}'_1 J\Theta \left(|\vec{v}_{CM}+\vec{v}'_1| - v_{\text{esc}}(r)\right)f_{\eta}\left(\sqrt{(\vec{v}_{CM}+\vec{v}_1')^2-v^2_{\text{esc}}(r)}\right)\times\nonumber \\ 
&\times f_i\left(\vec{v}_{CM}+\vec{v}_n^{CM}\right) \left(1+\frac{M_{\text{DM}}}{m_i}\right)|\vec{v'}_1|\sigma_{c,i}.
\end{align}
Without loss of generality, we orient the $z$-axis of our frame along $\vec{v}'_1$, so that $d^3\vec{v}'_1 = 4\pi v^{'2}_1dv'_1$. We let $\alpha$ and $\phi_{CM}$ specify the polar and azimuthal angles of $\vec{v}_{CM}$ with respect to the $z$-axis. Next we introduce the following definition to ease notation:  
\begin{align}
\beta(r)\equiv \frac{v^2_{\text{esc}}(r)-v_{CM}^2-v_1^{'2}}{2v_{CM}v'_1}.
\label{defn:beta}
\end{align}
Using the definitions above we write the capture rate per unit volume as an integral over four kinematic variables, $(v_{CM}, v'_1, c\alpha , \phi_{CM})$, where for brevity we write $c\alpha\equiv\cos\alpha$ and $s\alpha\equiv\sin\alpha$,
\begin{align}
\frac{dC_{\text{c}, i}}{dV} &= n_i(r)n_{\text{DM}}\int v^2_{CM}dv_{CM}dc\alpha d\phi_{CM}\int v^{'2}_1dv'_1 4\pi \Theta (c\alpha-\beta(r) )\times\nonumber \\ 
&\times f_{\eta}\left(\sqrt{(\vec{v}_{CM}+\vec{v}_1')^2-v^2_{\text{esc}}(r)}\right) f_i\left(\vec{v}_{CM}+\vec{v}_n^{CM}\right) \left(1+\frac{M_{\text{DM}}}{m_i}\right)^4v'_1\sigma_{c,i}.
\label{Eqn:dCdVAux1}
\end{align}
The two Maxwell distributions in the CM frame read
\begin{align}
&f_i\left(\vec{v}_{CM}+\vec{v}_n^{CM}\right) = \left(\frac{m_i}{2\pi  T_{\text{Sun}}}\right)^{3/2}& \exp{\left(-\frac{m_i}{2T_{\text{Sun}}}\left(v^2_{CM} +\frac{M_{\text{DM}}^2}{m_i^2}v_1^{'2} -2\frac{M_{\text{DM}}}{m_i}v_{CM}v_1'c\alpha\right)\right)} 
\end{align}
for the nuclei, and
\begin{align}
f_{\eta}\left(\sqrt{(\vec{v}_{CM}+\vec{v}_1')^2-v^2_{\text{esc}}(r)}\right) &= \frac{1}{\pi^{3/2}}\frac{e^{-\eta^2}}{2\eta}\frac{3}{2\bar{v}^2}\frac{\exp{\left(-\frac{3}{2\bar{v}^2} 2v_{CM}v'_1(c\alpha-\beta(r))\right)}}{\sqrt{2v_{CM}v'_1(c\alpha-\beta(r)) }} \times \nonumber \\&\times\sinh{\left(\sqrt{\frac{3}{2\bar{v}^2}}2\eta\sqrt{ 2v_{CM}v'_1(c\alpha-\beta(r)) }\right)}
\end{align}
for the DM particles, where $\eta^2\equiv 3\tilde{v}^2/(2\bar{v}^2)$.
\par
Next we consider the calculation of the capture cross-section $\sigma_{c, i}$. Similar to the zero-temperature calculation we will consider coherent nuclear scattering, such that the nuclear cross-section is given in terms of the nucleon cross-section, Eq.~\ref{Eqn:sigmaNuclear}, with the matrix element squared again given by Eq.~\ref{Eqn:MSqCc}.   The major difference in the finite temperature calculation is that the velocities  $(\vec{v}_{CM}, \vec{v}_1', \vec{v}_{1,f}')$ are no longer coplanar in general.  The argument of the kinematic step function that enforces the requirement that the final state DM particle is gravitationally bound is therefore significantly more involved. 
After some algebraic manipulations of Eq.~\ref{Eqn:sigma_c}, the capture cross-section can be written
\begin{align}
\sigma_c = \frac{\bar{m}^2}{M_{\text{DM}}^4}\frac{\alpha\alpha_D\epsilon^2}{v_1^{'4}}\int d\cos\theta d\phi_{\chi}\frac{1}{(1-\cos\theta +\mu)^2}\Theta (\beta(r) - \cos\gamma),
\label{egn:sigmac}
\end{align}
where $\gamma$ is the angle between $\vec{v}_{CM}$ and the final state DM velocity in the CM frame, $\vec{v}_{1,f}'$:
\begin{align}
\cos\gamma = \sin\alpha\sin\theta\cos (\phi_{\chi}-\phi_{CM}) + \cos\alpha\cos\theta.
\label{defn:gamma}
\end{align}
Here we can see that the singularity structure is (as expected for cross-sections with forward singularities) dependent on the integral over the polar scattering angle, while the criteria for capture make the integration in Eq.~\ref{egn:sigmac} over the azimuthal angle $\phi_\chi$ nontrivial.  As forward scattering cannot be neglected for long-range capture, the simpler average treatment of \cite{Busoni:2017mhe} is insufficient to capture the parametric dependence of the capture cross-section in this regime.  

\begin{figure}[t]
\begin{subfigure}{0.526\textwidth}
  \vspace{-0.22cm}
  \hspace{-0.32cm}
  \includegraphics[width=1.0\linewidth]{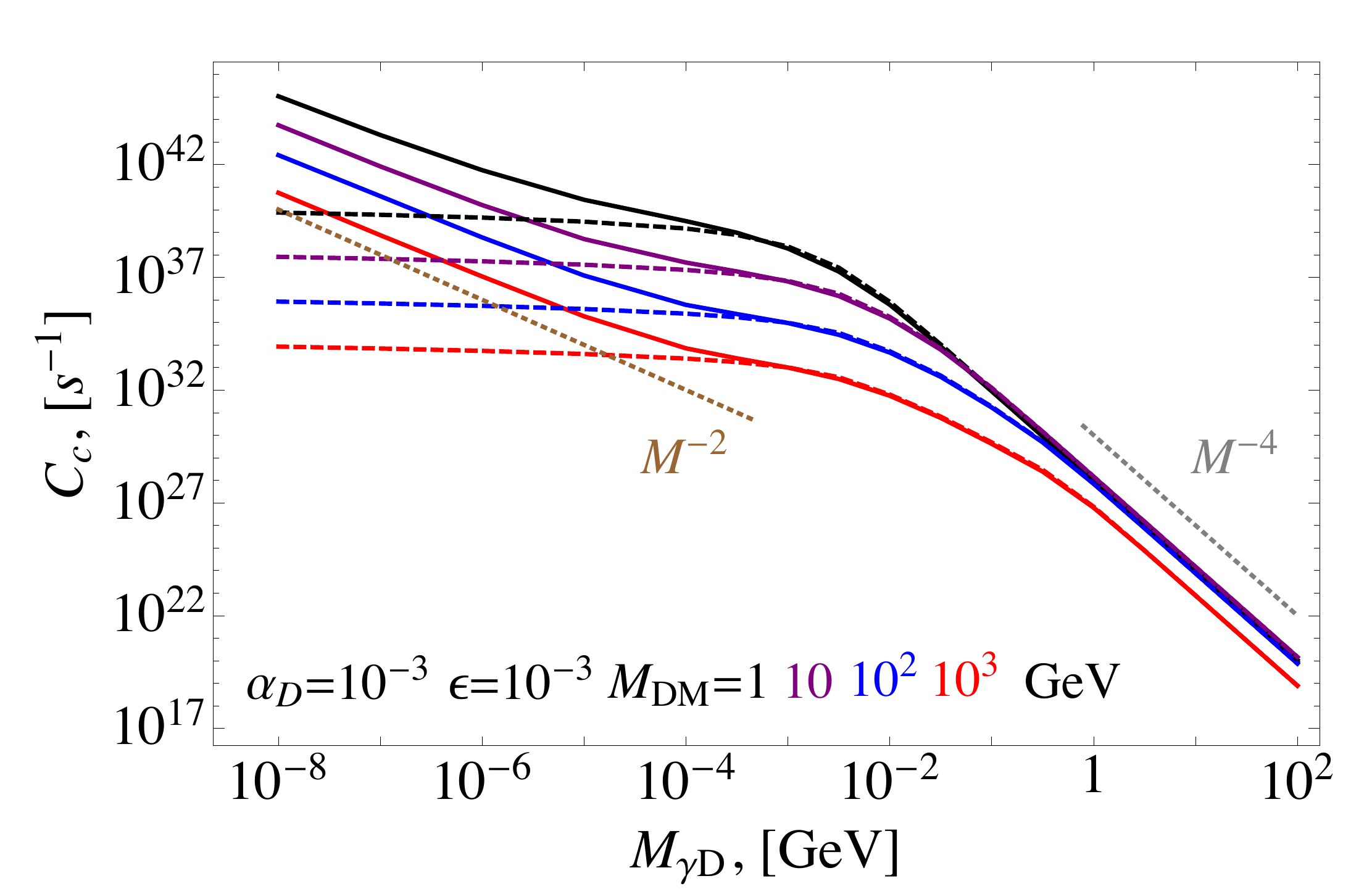}
\end{subfigure}%
\begin{subfigure}{0.5\textwidth}
  \hspace{-0.5cm}
  \includegraphics[width=1.0\linewidth]{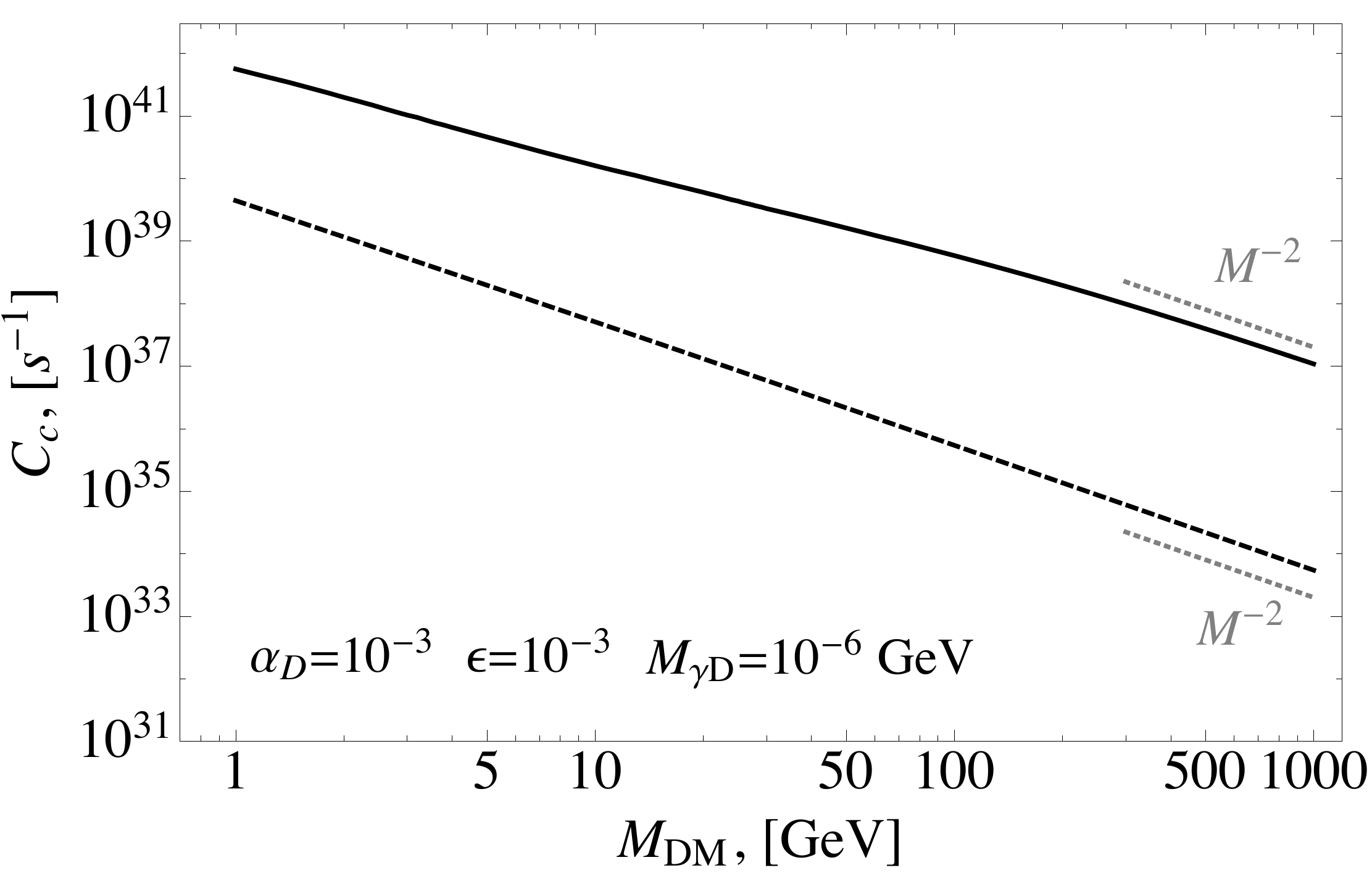}
\end{subfigure}
\caption{{\bf Left}: Solar capture rate as a function of $M_{\gamma_D}$ with (solid) and without (dashed) nuclear thermal motion. The different colors correspond to different DM masses. Dotted lines are included to highlight the parametric  dependence on $M_{\gamma_D}$ in the long and short-range regimes. {\bf Right}:  Solar capture rate as a function of $M_{\text{DM}}$ with (solid) and without (dashed) nuclear thermal motion for fixed $M_{\gamma_D}=10^{-6}$ GeV. 
We fix $\alpha_D=10^{-3}$, $\epsilon=10^{-3}$ and a uniform temperature $T_{\text{Sun}}=1.57\times 10^{7}$ K.}
\label{plot:Ccapture}
\end{figure}

Before we put all the calculations together, let us introduce one more definition that collects together three of the four angular integrals:
\begin{align}
J_c\equiv\frac{1}{4\pi}\int_0^{2\pi}d\phi_{CM}\int_{-1}^{1} d\cos\theta \int_0^{2\pi} d\phi_{\chi}\frac{1}{(1-\cos\theta +\mu)^2}\Theta (\beta(r) - \cos\gamma).
\label{Eqn:Jc}
\end{align}
The step functions in Eqs.~\ref{Eqn:dCdVAux1} and~\ref{egn:sigmac} require $-1\leq\cos\gamma\leq\beta(r)\leq c\alpha\leq 1$. We also note the reduced functional dependence $J_c(M_{\text{DM}}, M_{\gamma_D}, r, v_{CM}, v'_1, c\alpha) = J_c(\beta(r), c\alpha, \mu)$. We exploit this property to precompute $J_c$ numerically, before computing the remaining kinematic integrals over $v_{CM}, v'_1$ and $c\alpha$. We delegate the details of this calculation to Appendix~\ref{appendix.A}. Putting everything together, the capture rate per unit volume reads
\begin{align}
\frac{dC_{\text{c}, i}}{dV} &= n_i(r)n_{\text{DM}} (4\pi)^2 Z_i^2 \frac{\alpha\alpha_D\epsilon^2}{\bar{m}^2_i}\int_0^{\infty}dv_{CM}\int_0^{\infty}dv'_1\int dc\alpha\Theta(c\alpha -\beta(r))\Theta(1-\beta(r))\times \nonumber \\
&\times \Theta(\beta(r)+1) \frac{v^2_{CM}}{v'_1} f_i\left(v_{CM}, v'_1, c\alpha\right)f_{\eta}\left(v_{CM}, v'_1, c\alpha\right)J_c(\beta(r), c\alpha, \mu).
\end{align}
The kinematic constraint $-1\leq \beta(r) \leq 1$ translates into bounds on the speeds $v_{CM}$ and $v_1'$:
\begin{align}
v_{\text{esc}}(r) \leq v_{CM}+v_1' \quad \text{and} \quad v_{\text{esc}}(r) \geq |v_{CM} - v_1'|.
\end{align}
Thus, we finally obtain the following capture rate per volume shell:
\begin{align}
\frac{dC_{\text{c}, i}}{dV} &= n_i(r)n_{\text{DM}} (4\pi)^2 Z_i^2 \frac{\alpha\alpha_D\epsilon^2}{\bar{m}^2_i} \int_0^{\infty} dv_{CM}\int_{|v_{\text{esc}}(r)-v_{CM}|}^{v_{\text{esc}}(r)+v_{CM}}dv'_1\times  \nonumber\\
&\times\int dc\alpha\Theta(c\alpha -\beta(r))\frac{v^2_{CM}}{v'_1} f_i\left(v_{CM}, v'_1, c\alpha\right)f_{\eta}\left(v_{CM}, v'_1, c\alpha\right)J_c(\beta(r), c\alpha, \mu).
\end{align}
Together with the precomputed $J_c$, we numerically calculate the capture rate
\begin{align}
C_{\text{c}} = \sum_i C_{\text{c}, i} = \sum_i \int_0^R dr 4\pi r^2\frac{dC_{\text{c}, i}}{dV}.
\end{align}
 The resulting capture coefficient is shown in Fig.~\ref{plot:Ccapture} as a function of $M_{\text{DM}}$ and $M_{\gamma_D}$. 

\par
We again note the presence of two qualitatively distinct scattering regimes, corresponding to different scalings of $C_{\text{c}}$ with $M_{\gamma_D}$. For  mediator masses large compared to typical momentum transfers, we again obtain the $M_{\gamma_D}^{-4}$ scaling indicative of an effectively pointlike interaction, similar to the $T=0$ calculation. Numerically, we find that for a contact interaction the full thermal calculation gives a capture coefficient which is $\mathcal{O}(10\%)$ smaller than the $T=0$ calculation, in agreement with previous studies~\cite{Busoni:2013kaa}.  This small decrease in the capture coefficient comes from the small increase in the typical CM energy of the collision, which reduces the volume of phase space that meets the kinematic criteria for capture.
\par
Qualitatively new behavior occurs for small mediator masses, where we find that $C_\text{c}$ is proportional to $M_{\gamma_D}^{-2}$. This parametric dependence on the mediator mass is much stronger than the logarithmic sensitivity obtained in the $T=0$ calculation and leads to a much larger capture coefficient in the long-range regime. In the presence of target thermal motion, the subspace of kinematic configurations that can lead to capture at small momentum transfer is larger than the same subspace when thermal motion is absent. In the $T=0$ calculation, the logarithmic scaling with $M_{\gamma_D}$ is a consequence of the fact that capture with zero-momentum transfer can occur only at one kinematic point, $w=v_{\text{esc}}(r)$, for the incoming DM velocity. In the presence of thermal motion of the targets, capture via zero-momentum transfer can occur in the larger kinematic region $w\in [v_{\text{esc}}(r), v_{\text{esc}}(r)+\mathcal{O}(\sqrt{3T_{\text{Sun}}/m_i})]$. The forward singularity is now relevant in a finite (even if small) slice of the kinematic integral rather than at a single point, thus promoting the would-be logarithmic dependence on $\mu$ to a linear one. In turn, this leads to the capture rate scaling as $\mu^{-1} \sim M^{-2}_{\gamma_D}$. 
\par
One important corollary of the discussion above is that the capture rate calculation at finite $T$ remains insensitive to the high speed tail of the halo DM distribution. Long-range gravitational capture is dominated by scattering with small momentum transfer, which involves DM particles from the small velocity tail of the halo distribution. We have checked numerically that cutting the galactic frame DM velocity distribution off at the galactic escape speed  $v_{gal} \in [450\text{ km/s}, 650 \text{ km/s}]$~\cite{Baratella:2013fya} results in negligible corrections to the capture rate.  The dominance of small momentum transfer events in the long-range capture rate also means that DM-nuclear scattering will typically be well-approximated as coherent.

Finally, while we have used a constant core temperature for simplicity, the Sun's temperature varies with radius.  Since most capture occurs near the core, this is a reasonable first approximation.  As the thermal velocity of the target nuclei is important for controlling the thermal enhancement to the long-range capture rate, we expect that accounting for the radial variation of the solar temperature is more important for long-range capture than short-range capture.  We have checked numerically that incorporating the radial temperature profile in the model of Ref.~\cite{Bahcall:2004pz} reduces the finite-temperature capture rate by up to tens of percent in the long-range regime, which can be important for detailed studies of signals in any specific model realizing long-range nuclear interactions.


\section{Dark Matter self-capture and self-ejection with thermal corrections}
\label{s.selfcaptAndselfejec}

In this section we discuss DM self-capture and self-ejection rates. Self-capture occurs when an incoming DM particle scatters against a previously-captured DM particle, such that both particles remain gravitationally bound after the collision. Conversely, self-ejection occurs when both DM particles are gravitationally unbound after scattering.  The self-capture coefficient $C_\mathrm{sc}$  gives the rate of gravitational self-capture per bound DM particle, and similarly $C_\mathrm{se}$ for self-ejection.  In our reference DM model, the DM particle is Dirac; we assume the relic abundance of $\chi$ and $\bar\chi$ are equal.  The calculation of $C_\mathrm{sc}$ proceeds similarly to that for nuclear capture, except that the target particle is now a previously-captured DM particle instead of a nucleus.  The cross-section for events leading to the net capture of DM is therefore subject to the additional kinematic constraint that the target DM particle remain bound after the collision; similarly, for self-ejection, an additional kinematic requirement imposes that the target DM particle is unbound following the collision.

To compute the rate for either process, one must model the distribution of the target DM population inside the capturing object. We consider the case where the DM-SM interaction is strong enough to thermalize the captured DM particles with the nuclei within the lifetime of the Sun.\footnote{See Ref.~\cite{Gaidau:2018yws} for a discussion of thermalization via short-range spin- and velocity-independent nuclear interactions.  The dark photon-mediated reference model used here yields spin- and velocity-independent nuclear interactions, but depending on the dark photon mass and the momentum transfer can produce either short- or long-range interactions.  The approach to thermalization involves multiple scatterings with increasingly small momentum transfer; thus even if the initial capture collision is in the long-range regime, the process of thermalization may be dominated by short-range scatterings.}  The spatial distribution of the DM is then given by the equilibrium quantity
\begin{align}
n_c(r) &= A\exp\left(-M_{\text{DM}}v^2_{\text{esc}}(r)/2T_{\text{Sun}}\right),
\label{eqn:Csc1}
\end{align}
which for simplicity we again take at the core temperature of the capturing body.  Here the normalization factor  $A$ ensures  $\int dV n_c(r) =1$. We will again take the Sun's temperature to be constant for simplicity.  For self-capture of heavy DM ($M_{\text{DM}}\gtrsim 10$ GeV), unlike nuclear capture, this is a good approximation even in the long-range regime simply  because the physical volume occupied by the captured dark matter population is small,  occupying a sphere with radius at most $\mathcal{O}(2-3\%)$ of $R_\odot$, where the variation of the solar temperature is small.

\subsection{Zero-temperature calculation}

We begin with a review of the self-capture calculation in the limit of no thermal motion of the captured DM particles~\cite{Zentner:2009is,Fan:2013bea}. The self-capture coefficient is constructed in analogy to the nuclear capture coefficient, by integrating contributions from all volume shells:
\begin{align}
\frac{dC_{\text{sc}}}{dV} = \frac{1}{2}n_{\text{DM}}n_c(r)\int_{0}^{\infinity}d^3\vec{w}\Theta(|\vec{w}|-v_{\text{esc}}(r))f_{\eta}\left(\sqrt{w^2 - v_{\text{esc}}^2(r)}\right)|\vec{w}|\sigma_{sc} .
\end{align} 
The differences with respect to Eq.~\ref{eqn:Cc320} for the nuclear capture coefficient are (i) the target particle distribution is now given by $n_c(r)$; (ii) an overall factor of 1/2, stemming from the assumption that the relic abundance of DM is symmetric; (iii) the relevant cross-section is now that for self-capture, $\sigma_{sc}$.

 At the microscopic level several processes contribute to $\sigma_{sc}$: self-capture of incoming particles (antiparticles) by scattering against captured particles (antiparticles), and of particles (antiparticles) by scattering on target antiparticles (particles). In the non-relativistic limit the matrix element squared, summing all contributions, can be written
\begin{align}
&|\mathcal{M}|^2 = \frac{4g^4_D}{v^{'4}_1}f_{sc}(\cos\theta, \mu), 
\label{eqn:MatrixElementSC}
\end{align} 
with
\begin{align}
f_{sc}(\cos\theta, \mu) \equiv \frac{3}{(1-\cos\theta+\mu)^2}&+\frac{1}{(1+\cos\theta+\mu)^2}+\frac{1}{(1-\cos\theta+\mu)}\frac{1}{(1+\cos\theta+\mu)},
\label{defn:fsc}
\end{align} 
where $\theta$ is the scattering angle in the CM frame and $\mu$ is defined in Eq.~\ref{Eqn:mu_defn}. The self-capture condition demands
\begin{align}
|\vec{v}_{1, f}^{lab}| \leq v_{\text{esc}}(r) \quad \text{ and } \quad |\vec{v}_{2, f}^{lab}| \leq v_{\text{esc}}(r).
\label{SelfCaptureCondition}
\end{align}
The Galilean transformation from the CM frame to the lab frame reads
\begin{align}
\vec{w} = \vec{v}'_1 + \vec{v}_{CM},  \phantom{spsp}
\vec{v}_c = 0 = -\vec{v}'_{1} + \vec{v}_{CM}
\end{align}
where $\vec{v}_c = 0$ is a consequence of neglecting the thermal motion of the target DM particles. Now we write the self-capture condition as a constraint on the CM frame scattering angle $\theta$:
\begin{align}
\Theta_{sc} \equiv \Theta(\beta(r) -|\cos\theta|),
\label{Theta_sc}
\end{align}
noting that $\beta\geq 0$ is a necessary condition for self-capture to occur. Furthermore, Eq.~\ref{defn:beta} simplifies to $\beta(r) = 2v^2_{\text{esc}}(r)/w^2 -1$, ensuring that $\beta\leq 1$. The self-capture cross-section is found by integrating the matrix element with $\Theta_{sc}$ specified in Eq.~\ref{Theta_sc}, over the solid angle in the CM frame:
\begin{align}
\sigma_{sc}= \frac{1}{256\pi^2M^2_{\text{DM}}}\int d\Omega_{f, \chi}|\mathcal{M}|^2\Theta_{sc} = \frac{4\pi\alpha^2_D}{M^2_{\text{DM}}w^4}\Theta(\beta(r))\int_{-\beta(r)}^{\beta(r)}d\cos\theta f_{sc}(\cos\theta, \mu). 
\label{eqn:sigmaSCAux1}
\end{align}
The self-capture cross-section $\sigma_{sc}$ can be evaluated in closed form,
\begin{align}
\sigma_{sc} = \frac{4\pi \alpha^2_D}{M^2_{\text{DM}}w^4}\Theta (\beta(r))\left[\frac{8\beta(r)}{-\beta^2(r)+(1+\mu)^2}+\frac{2\text{ArcTanh}\left(\frac{\beta(r)}{1+\mu}\right)}{1+\mu}\right].
\end{align}
Collecting terms, we obtain the following expression for self-capture per unit volume:
\begin{align}
\frac{dC_{\text{sc}}}{dV} =\frac{1}{2}n_{\text{DM}} n_c(r) & \int_{v_{\text{esc}}(r)}^{\sqrt{2}v_{\text{esc}}(r)}\frac{dw}{w} 4\pi f_{\eta}\left(\sqrt{w^2 - v_{\text{esc}}^2(r)}\right) \times \nonumber \\ 
&\times \frac{4\pi \alpha^2_D}{M^2_{\text{DM}}}\left[\frac{8\beta(r)}{-\beta^2(r)+(1+\mu)^2}+\frac{2\text{ArcTanh}\left(\frac{\beta(r)}{1+\mu}\right)}{1+\mu}\right].
\end{align}
The term in the square brackets is singular at $w=v_{\text{esc}}(r)$ for $\mu=0$, resulting in a logarithmic dependence of $C_{\text{sc}}$ on $\mu$, and in turn on the dark photon mass $M_{\gamma_D}$, similar to zero-temperature nuclear capture. The characteristic scattering regime is determined by how large the mediator mass is compared to the typical momentum transfer, $M_{\text{DM}}v_{\text{esc}}\sim (2\times10^{-3})M_{\text{DM}}$: heavy mediators correspond to pointlike interactions, giving rise to the expected $M^{-4}_{\gamma_D}$ dependence;  in the opposite limit, the dependence is logarithmic.

\subsection{Self-Capture at Finite Temperature}

Now we turn to self-capture at finite temperature. 
We model the velocity distribution of captured DM particles with a  Maxwell distribution at the core temperature, truncated at the local escape velocity:\begin{align}
f_c(\vec{v}_c) &= A_{\text{MB}}(r)\left(\frac{M_{\text{DM}}}{2\pi T_{\text{Sun}}}\right)^{3/2}\exp{\left(-\frac{M_{\text{DM}}v_c^2}{2T_{\text{Sun}}}\right)},
\qquad \qquad |\vec{v}_c| < v_{\text{esc}}(r),
\label{eqn:DMdistroMB}
\end{align}
with $A_{\text{MB}}(r)$ chosen such that $\int_0^{v_{\text{esc}}(r)}d^3\vec{v}_cf_c(\vec{v}_c)=1$.
The self-capture coefficient now includes an integral over the initial velocity of the captured DM particle, $\vec{v}_c$:
\begin{align}
\frac{dC_{\text{sc}}}{dV} &= \frac{1}{2}n_{\text{DM}}n_c(r)\int_{0}^{\infinity}d^3\vec{u} \frac{w}{u}f_{\eta}(u)\int_0^{v_{\text{esc}}(r)} d^3\vec{v}_cf_c(v_c)|\vec{w}-\vec{v}_c|\sigma_{sc}.
\label{eqn:SCcoefficientAux3}
\end{align}

The kinematic constraints leading to self-capture, given in Eq.~\ref{SelfCaptureCondition}, now translate into the following step function on the scattering angle $\gamma$, defined in Eq.~\ref{defn:gamma}:
\begin{align}
\Theta_{sc} \equiv \Theta_{sc} (\beta(r) - |\cos\gamma|)\times \Theta (\beta(r)).
\label{eqn:ThermalThetaSC}
\end{align}

We make a Galilean transformation to bring the velocity integrals to the CM frame, where $\vec{v}'_1$ and $\vec{v}'_2$ denote the CM velocities of the incoming and captured DM particles before the collision,
\begin{align}
&\int_{0}^{\infinity}d^3\vec{u} \frac{w}{u}f_{\eta}(u)\int_0^{v_{\text{esc}}(r)} d^3\vec{v}_cf_c(v_c) = \int_{v_{\text{esc}}(r)}^{\infinity}d^3\vec{w}f_{\eta}\left(\sqrt{w^2-v_{\text{esc}}^2(r)}\right)\int_0^{v_{\text{esc}}(r)} d^3\vec{v}_cf_c(v_c) =  \nonumber\\
&=\int_{0}^{\infinity}d^3\vec{w}f_{\eta}\left(\sqrt{w^2-v_{\text{esc}}^2(r)}\right)\Theta (|\vec{w}|-v_{\text{esc}}(r))\int_0^{\infinity} d^3\vec{v}_cf_c(v_c)\Theta (v_{\text{esc}}(r)-|\vec{v}_c|).
\end{align}
Now we switch to CM frame velocities,
\begin{align}
\vec{w} = \vec{v}'_1 + \vec{v}_{CM}, \phantom{spsp}
\vec{v}_c = \vec{v}'_{2} + \vec{v}_{CM}
\label{CM:vel}
\end{align}
with $\vec{v}'_{2}=-\vec{v}'_{1}$. The Jacobian of the transformation $(\vec{w}, \vec{v}_{c})\rightarrow (\vec{v}_{CM} , \vec{v}'_1)$  is now $J=8$. In the CM frame, the velocity integrals read
\begin{align}
&\int_0^{\infinity}d^3\vec{v}_{CM}\int_0^{\infinity}d^3\vec{v}'_1Jf_{\eta}\left(\sqrt{(\vec{v}'_1+\vec{v}_{CM})^2-v_{\text{esc}}^2(r)}\right)f_c(|-\vec{v}'_1+\vec{v}_{CM}|)\times  \nonumber\\
&\times \Theta \left(|\vec{v}'_1+\vec{v}_{CM}| -v_{\text{esc}}(r)\right)\Theta \left(v_{\text{esc}}(r)-|-\vec{v}'_1+\vec{v}_{CM}|\right).
\end{align}
We choose coordinates where $\vec{v}'_1$ is along the $z$-axis, so that the solid angle integration for $\vec{v}'_1$ is trivial. We denote the polar and azimuthal angles of $\vec{v}_{CM}$ as $(\alpha , \phi_{CM})$. Using the definition of $\beta$ in Eq.~\ref{defn:beta} we write the velocity integrals in a more compact form,
\begin{align}
&\int_0^{\infinity}v^2_{CM}dv_{CM}\int_{-1}^{1}dc\alpha \int_0^{2\pi}d\phi_{CM}\int_0^{\infinity}4\pi v^{'2}_1dv'_1Jf_{\eta}\left(\sqrt{2v_{CM}v'_1(c\alpha-\beta(r))}\right)\times \nonumber \\ 
&\times f_c\left(\sqrt{v^2_{CM}+v^{'2}_1+2v_{CM}v'_1c\alpha}\right) \Theta(c\alpha - \beta(r))\Theta (c\alpha+\beta(r)).
\label{eqn:CMvelIntegralsSC}
\end{align}

\begin{figure}[t]
\begin{subfigure}{0.51\textwidth}
  \vspace{-0.11cm} 
  \hspace{-0.26cm}
  \includegraphics[width=1.0\linewidth]{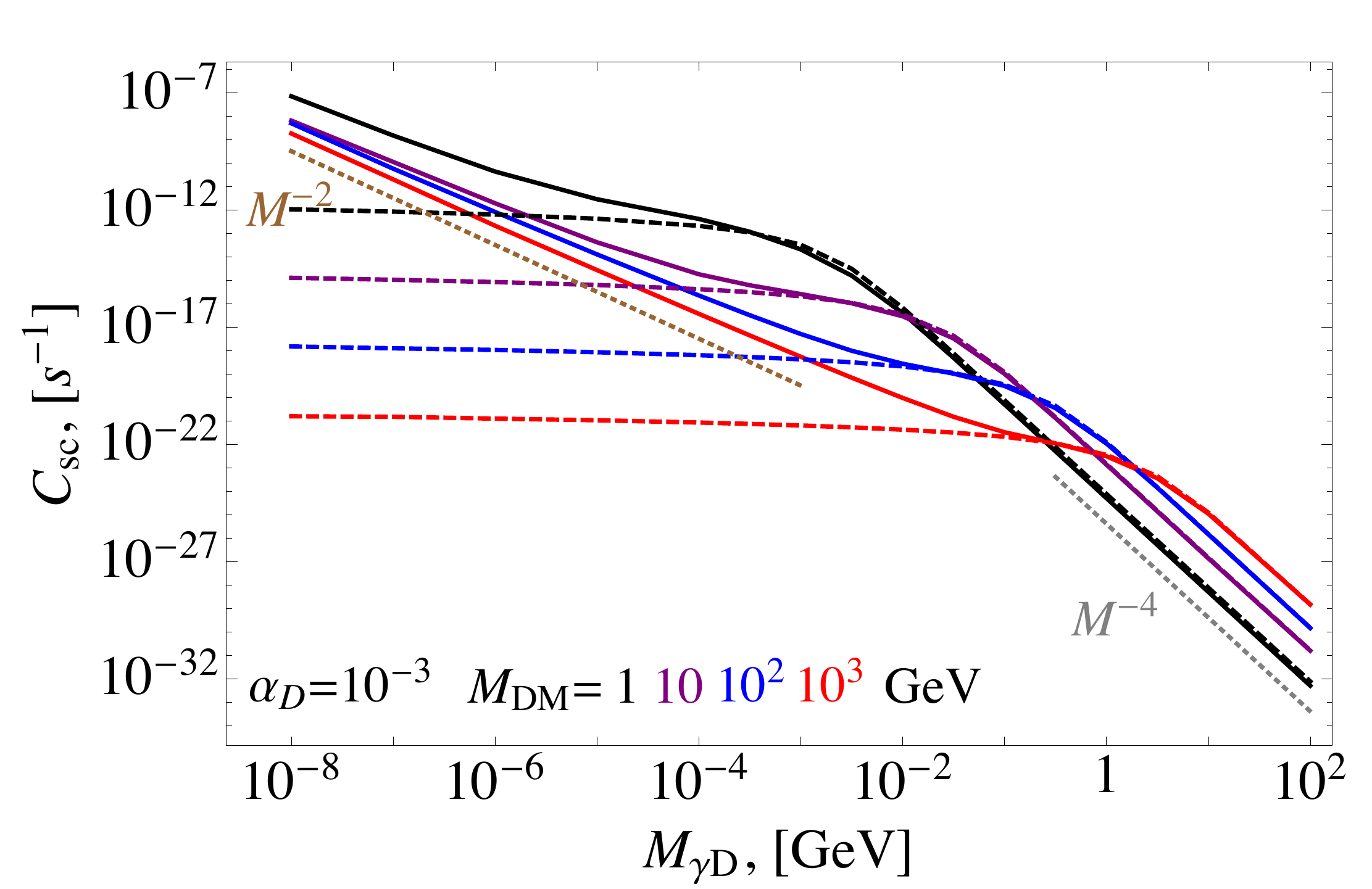}
\end{subfigure}%
\begin{subfigure}{0.498\textwidth}
  \vspace{0.0cm}
  \hspace{-0.26cm}
  \includegraphics[width=1.0\linewidth]{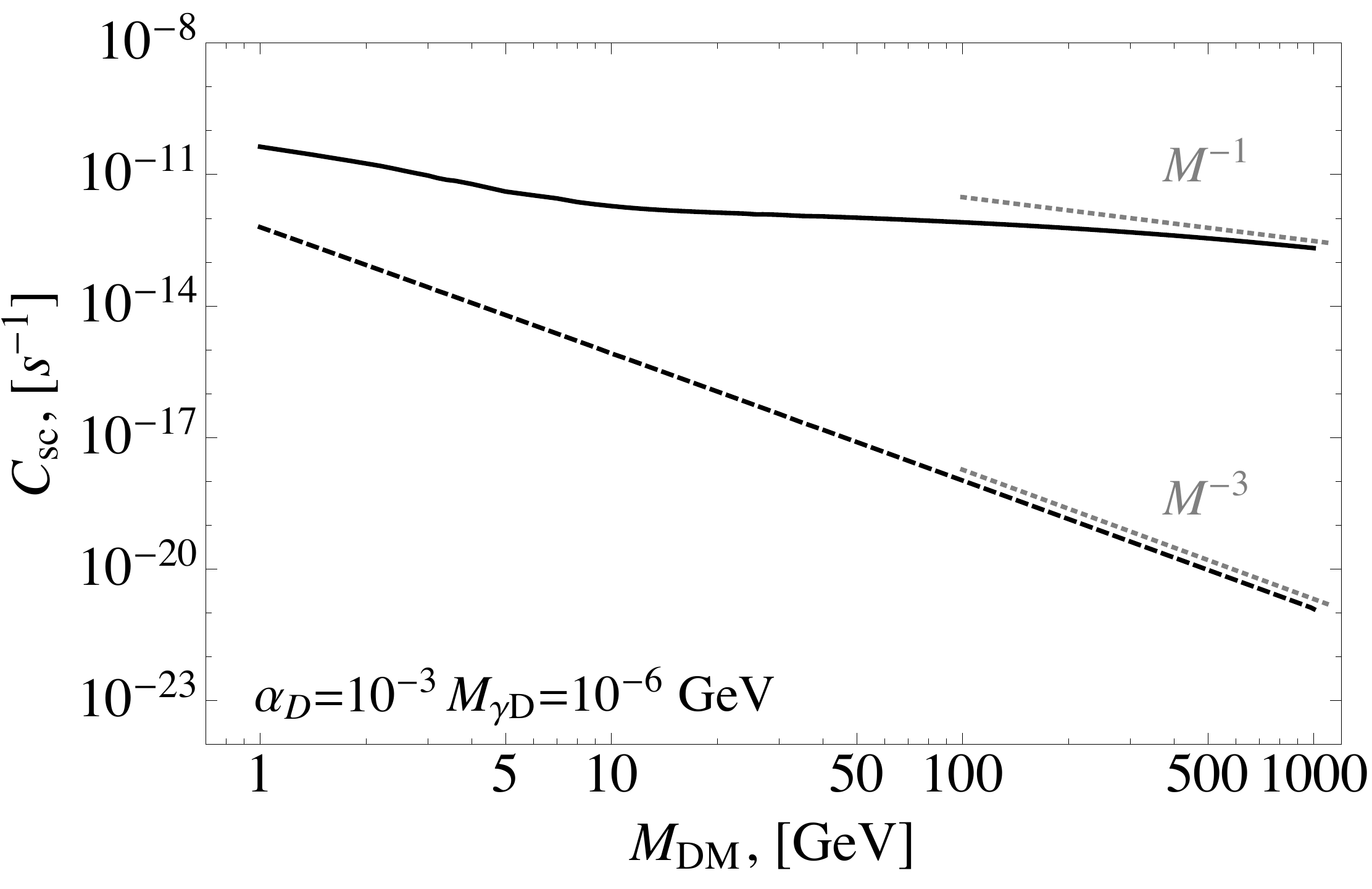}
\end{subfigure}
\caption{{\bf Left}: Self-capture coefficient in the Sun as a function of mediator mass $M_{\gamma_D}$. Solid lines correspond to the full calculation including the target's thermal motion, while dashed lines show the zero-temperature calculation. The different colors correspond to different DM masses. Dotted lines are included to highlight the parametric scaling with $M_{\gamma_D}$ in the two regimes. {\bf Right:} Self-capture coefficient in the Sun, as a function of DM mass. We fix $M_{\gamma_D}=10^{-6}$ GeV, corresponding to long-range scattering, and $\alpha_D=10^{-3}$.}
\label{plot:SunEarthCsc}
\end{figure}

\noindent
We calculate the self-capture cross-section $\sigma_{sc}$ in a manner similar to Eq.~\ref{eqn:sigmaSCAux1}, where the matrix element given by Eq.~\ref{eqn:MatrixElementSC} is convoluted with the new self-capture condition in Eq.~\ref{eqn:ThermalThetaSC},
\begin{align}
\sigma_{sc} &= \frac{1}{256\pi^2M^2_{\text{DM}}}\int d\cos\theta \,d\phi_{\chi}|\mathcal{M}|^2\Theta_{sc}=  \nonumber \\ 
&=\frac{\alpha^2_D}{4M^2_{\text{DM}}(v_1')^4}\Theta (\beta(r))\int  d\cos\theta d\phi_{\chi}f_{sc}(\cos\theta, \mu)\Theta (\beta(r)- |\cos\gamma|).
\end{align}
Putting together this cross-section expression with the CM-frame velocity integrals of Eq.~\ref{eqn:CMvelIntegralsSC} in Eq.~\ref{eqn:SCcoefficientAux3} gives for the self-capture coefficient 
\begin{align}
&\frac{dC_{\text{sc}}}{dV} = \frac{1}{2}n_{\text{DM}}n_{c}(r)\int_0^{\infinity}v^2_{CM}dv_{CM} \int_0^{\infinity}v^{'2}_1dv'_1\int_{-1}^{1}dc\alpha (4\pi)^2 f_{\eta}\left(\sqrt{2v_{CM}v'_1(c\alpha-\beta(r))}\right)\times \nonumber \\ 
&\times f_c\left(\sqrt{v^2_{CM}+v^{'2}_1+2v_{CM}v'_1c\alpha}\right) \Theta(c\alpha - \beta(r))\Theta (c\alpha+\beta(r))\Theta (\beta(r))\frac{2\alpha^2_D}{M^2_{\text{DM}}v_1^{'3}}J_{sc}(\beta(r), c\alpha, \mu),
\label{eqn:Cc313}
\end{align}
where we introduced the angular integral $J_{sc}$, 
\begin{align}
J_{sc} \equiv \frac{1}{4\pi}\int_0^{2\pi}d\phi_{CM}\int_0^{2\pi}d\phi_{\chi}\int_{-1}^{1}d\cos\theta f_{sc}(\cos\theta, \mu )\Theta (\beta(r)-|\cos\gamma |).
\label{defn:Jsc}
\end{align}
We detail the calculation of $J_{sc}$ in Appendix~\ref{appendix.B}. The three step functions in Eq.~\ref{eqn:Cc313} restrict the integration domain as follows. The first and third step functions make the second step function redundant. Additionally, the first step function implicitly demands that $\beta(r)\leq 1$. Hence, we must have $0\leq\beta(r)\leq 1$. Using the definition of $\beta$, this condition turns into a finite integration region for the speeds:
\begin{align}
\frac{dC_{\text{sc}}}{dV}& = \frac{1}{2}n_{\text{DM}}n_{c}(r)\int_0^{v_{\text{esc}}(r)}v^2_{CM}dv_{CM} \int_{v_{\text{esc}}(r)-v_{CM}}^{\sqrt{v^2_{\text{esc}}(r)-v^2_{CM}}}v^{'2}_1dv'_1\int_{\beta (r)}^{1}dc\alpha \times \nonumber \\ 
&\times f_{\eta}\left(\sqrt{2v_{CM}v'_1(c\alpha-\beta(r))}\right) f_c\left(\sqrt{v^2_{CM}+v^{'2}_1+2v_{CM}v'_1c\alpha}\right)2\frac{(4\pi)^2\alpha^2_D}{M^2_{\text{DM}}v_1^{'3}}J_{sc}(\beta(r), c\alpha, \mu).
\end{align}
In Fig.~\ref{plot:SunEarthCsc}  we present the  results of numerically evaluating the remaining integrals. 
%
%
The left panel shows the self-capture rate as a function of dark photon mass for several different DM masses.  The capture rate again exhibits two qualitatively distinct regimes: the short-range regime with $C_{\text{sc}}\propto M^{-4}_{\gamma_D}$ and the long-range regime where $C_{\text{sc}}\propto M^{-2}_{\gamma_D}$, similar to long-range nuclear capture. 
In the right panel, we plot the self-capture rate as a function of DM mass, contrasting the full calculation to the zero temperature one.  The dependence of  $C_{\text{sc}}$ on $M_{\text{DM}}$ comes from three sources: (i) the local number density of DM in the halo; (ii) the scattering cross-section, $\sigma_{sc}\propto M^{-2}_{\text{DM}}$; and the novel factor (iii) $M^{2}_{\text{DM}}/M^2_{\gamma_D}$ from the $\mu^{-1}$ scaling of the thermally-averaged self-capture cross-section.  Combining these factors produces a final $C^{T\neq 0}_{\text{sc}}\sim M^{-1}_{\text{DM}}$ scaling.  By contrast, in the zero-temperature calculation, the dependence on $\mu$ softens to a logarithmic one, and to leading order, $C^{T=0}_{\text{sc}}\sim M^{-3}_{\text{DM}}$.

\par
Finally, we discuss the self-ejection coefficient $C_{\text{se}}$. Most of the formalism and notation for this case carries over from  self-capture  and we only highlight the qualitative differences between the two processes. First, the self-ejection constraints on the final state speeds are reversed from those for self-capture,
\begin{align}
|\vec{v}^{lab}_{1,f}|\geq v_{\text{esc}}(r), \quad  \quad |\vec{v}^{lab}_{2,f}|\geq v_{\text{esc}}(r).
\end{align}
These conditions lead to the following step functions ensuring self-ejection:
\begin{align}
\Theta_{se} \equiv \Theta (-\beta(r)-|\cos\gamma|)\Theta (-\beta(r))\Theta (1+\beta(r)).
\end{align}
Recall that self-capture demands $0\leq\beta(r)\leq1$, while for self-ejection $-1\leq\beta(r)\leq 0$. We minimize the computational cost of the numerical integrals in this case by observing that the contribution to self-ejection at $\bar{\beta}\equiv-\beta$ is equal to the contribution to self-capture at $\beta$. This observation is a consequence of the fact that the same microscopic scattering processes are responsible for both self-capture and self-ejection, and the difference between the two processes is merely kinematic. Then,
\begin{align}
\sigma_{se}(\beta(r), c\alpha ,\mu) = \sigma_{sc}(\bar{\beta}(r), c\alpha , \mu). 
\end{align}
Therefore, we can construct the self-ejection coefficient per unit  volume  using Eq.~\ref{eqn:Cc313} by exploiting this symmetry:
\begin{align}
&\frac{dC_{\text{se}}}{dV} = \frac{1}{2}n_{\text{DM}}n_{c}(r)\int_0^{\infinity}v^2_{CM}dv_{CM} \int_0^{\infinity}v^{'2}_1dv'_1\int_{-1}^{1}dc\alpha \Theta(1-\bar{\beta}(r))\Theta (\bar{\beta}(r))  \Theta(c\alpha +\bar{\beta}(r)) \times \nonumber \\ 
&\times f_{\eta}\left(\sqrt{2v_{CM}v'_1(c\alpha+\bar{\beta}(r))}\right)f_c\left(\sqrt{v^2_{CM}+v^{'2}_1+2v_{CM}v'_1c\alpha}\right)2\frac{(4\pi)^2\alpha^2_D}{M^2_{\text{DM}}v_1^{'3}}J_{sc}(\bar{\beta}(r), c\alpha, \mu).
\end{align}
At zero temperature, self-ejection is negligible compared to self-capture in the Sun for  both long- and short-range interactions~\cite{Zentner:2009is, Fan:2013bea, Gaidau:2018yws}. We have verified that this remains true for $T\neq 0$.


\section{Implications for the Captured Dark Matter Population}
\label{s.DiscussionImplications}

As demonstrated in the two previous sections,  retaining the finite thermal motion of the target particles is critical in correctly evaluating capture rates through long-range interactions ($\mu \ll 1)$.   
To help understand the range of mediator masses where these thermal target effects are quantitatively significant for a given astrophysical object,
we begin by estimating the scale of momentum transfer in DM-nucleus collisions. Using Eqs.~\ref{GalTransform} and~\ref{vCMEqn}, the momentum exchange in a given scattering between halo DM particles and a nuclear target is 
\begin{align}
(\Delta \vec{p})^2 &= M^2_{\text{DM}}v_1^{'2} = M^2_{\text{DM}}|\vec{w}-\vec{v}_{CM}|^2 = \frac{ M^2_{\text{DM}}m_i^2}{( M_{\text{DM}}+m_i)^2}|\vec{w}-\vec{v}_n|^2.
\end{align}
Now, $v_n$ is drawn from a Maxwell distribution at the (core) temperature of the massive body, hence $v^2_n \sim 3T/m_i$. Meanwhile, $w^2 = u^2 +v^2_{\text{esc}}(r)$, where the natural scale for $u$ is $\bar{v}$, the rms speed of the halo particles. Hence, the typical momentum transfer in a scattering event can be estimated as
\begin{align}
\Delta p_{typ}= \bar{m}_i \sqrt{v^2_{\text{esc}}(r) +\bar{v}^2 +3T/m_i -2\sqrt{v^2_{\text{esc}}(r) +\bar{v}^2}\sqrt{3T/m_i}\cos(\measuredangle(\vec{w}, \vec{v}_n))},
\end{align}
up to factors of a few. In the Sun, the three velocity scales that control the kinematics are
\begin{align}
&\bar{v} = 288\text{ km/s}, \nonumber \\
&v_{\text{esc}}(r)\in [600, 1300] \text{ km/s}, \nonumber \\
&\sqrt{3T_{\text{Sun}}/m_i}\sim 85 \text{ km/s}\times \left(\frac{50 \text{ GeV}}{m_i}\right)^{1/2}.
\end{align}
Thus for scattering in the Sun, the dominant velocity scale is $v_{\text{esc}}$, giving
\begin{align}
\Delta p_{typ} \approx 5\times 10^{-3}\bar{m}_i.
\end{align}
For collisions with oxygen nuclei, $\Delta p_{typ}= 16$ MeV for heavy dark matter, $M_{\text{DM}}\gg 16$ GeV.  For  $M_{\text{DM}}\ll 16$ GeV,  $\Delta p_{typ}= 1\,\text{MeV}(M_{\text{DM}}/\text{GeV})$.  Thus for gravitational capture in the Sun, typical DM-nuclear collisions are effectively short-range for mediator masses heavier than the 10 MeV scale.    

Electrons have the same temperature as nuclei, but owing to their smaller mass will have a much larger thermal velocity. Thus for electrons the thermal velocity can easily dominate over the other velocity scales in evaluating $\Delta p_{typ}$. However, for DM heavier than the evaporation mass in the Sun, the reduced DM-electron mass is $\bar{m}_i\approx m_e$, which suppresses  $\Delta p_{typ}$ in comparison to nuclear scattering. We thus expect scattering on electrons to remain subleading in this DM mass range.

Meanwhile the typical momentum exchange in collisions between halo and bound DM, relevant for self-capture, follows by replacing the DM-nucleus reduced mass $\bar m_i$ with the reduced mass for the DM-DM system.  Thus while 
the typical momentum transfer in nuclear capture collisions saturates as the DM mass increases, in the sense that $(\Delta p_{typ})^{\text{c}}=\bar{m}_i v_{typ}\approx m_i v_{typ}$, the typical momentum transfer in self-capture collisions $(\Delta p_{typ})^{\text{sc}}=1/2 M_{\text{DM}}v_{typ}$ increases with DM mass. Consequently, for heavy DM, long-range self-capture rates can exhibit a parametric enhancement for mediator masses where nuclear capture rates do not. 

For mediator masses $M_{\gamma_D} \lesssim  \Delta p_{typ}$, the short-range approximation is no longer adequate to describe gravitational capture.  However, as we have seen, the biggest thermal enhancements to capture rates arise from the forward scattering regime, where the momentum transfers can be substantially smaller than $\Delta p_{typ}$.  To understand precisely how large the hierarchy $M_{\gamma_D} \ll  \Delta p_{typ}$ needs to be for thermal enhancements to be sizeable, we numerically evaluate the capture rates and find the resulting captured dark matter population.

In the regime where DM reaches thermal equilibrium with the nuclei of the Sun, the number of bound DM particles in the Sun, $N(t)$, can be obtained in terms of the nuclear and self-capture coefficients, together with the annihilation coefficient $C_{\text{ann}}$, by solving
\beq
\label{eq:diffyq}
\frac{dN}{dt} = C_{\text{c}} +  C_{\text{sc}} N - C_{\text{ann}} N^2 .
\eeq
Here for brevity we have omitted subdominant processes such as self-ejection and (self-)evaporation, which can be neglected for DM masses above $\sim 4$ GeV ($\sim 5$ GeV).
In the generic situation where DM annihilations proceed dominantly to light mediators, the annihilation coefficient $C_{\text{ann}}$ is given by 
\begin{align}
C_{\text{ann}} & =\frac{1}{2}\langle\sigma_{\text{ann}} v\rangle \int dV n^2_c(r) \\
 & = \frac{1}{2}\frac{\pi\alpha_D^2}{M^2_{\text{DM}}}\langle S_S\rangle \int dV n^2_c(r),
 \end{align}
where $\langle S_S\rangle$ is the thermally-averaged Sommerfeld enhancement, with $\langle S_S\rangle \propto \alpha_D M^{1/2}_{\text{DM}}$ \cite{Fan:2013bea}, and $n_c(r)$ is given by Eq.~\ref{eqn:Csc1}.  

The solution to Eq.~\ref{eq:diffyq} is  \cite{Zentner:2009is}
\begin{align}
N(t) = \frac{C_{\text{c}}\text{tanh}\left(\frac{t}{\xi}\right)}{\frac{1}{\xi} -\frac{C_{\text{sc}}}{2}\text{tanh}\left(\frac{t}{\xi}\right)},
\label{eq:DMCapturedPopulation}
\end{align}
in terms of the equilibration time scale
\beq
\label{eq:xi}
\xi^{-1}=\sqrt{C_{\text{c}}C_{\text{ann}}+C^2_{\text{sc}}/4}. 
\eeq
Parametrically, for heavy DM in the long-range regime, the nuclear capture rate scales as
\begin{align}
\label{eq:ccscaling}
C_{\text{c}}\propto \frac{\epsilon^2\alpha_D}{M^2_{\text{DM}}M^2_{\gamma_D}} ,
\end{align}
while for self-capture 
\begin{align}
C_{\text{sc}}\propto \frac{\alpha_D^2}{M_{\text{DM}}M^2_{\gamma_D}}.
\end{align}
Meanwhile, since $ \int dV n^2_c(r) \propto M^{3/2}_{\text{DM}}$, $C_{\text{ann}} \propto \alpha_D^3$.  Thus $C_{\text{c}}C_{\text{ann}}\propto \epsilon^2\alpha_D^4 M^{-2}_{\text{DM}}M^{-2}_{\gamma_D}$ and \\ $ C_{\text{sc}}^2\propto \alpha_D^4 M^{-2}_{\text{DM}}M^{-4}_{\gamma_D}$.  Dialing the dark gauge coupling does not affect the relative importance of nuclear capture versus self-capture, since both  $C_{\text{sc}}^2$ and $C_{\text{c}}C_{\text{ann}}$ have the same dependence on $\alpha_D$ \cite{Feng:2016ijc}.  However, since at finite temperature  $C_{\text{c}}C_{\text{ann}}$ and $C_{\text{sc}}^2$ have different parametric dependences on $ M_\text{DM}$ and $M_{\gamma_D}$, the importance of self-capture compared to nuclear capture increases as the hierarchy between the dark photon and DM mass is increased.

In the rest of this section we examine the finite-temperature DM population in two different regimes for the mass of the dark photon.

\subsection{Visibly decaying dark photons in the Sun and beyond}
\label{sec:vis}

\begin{figure}[t]
\begin{subfigure}{0.61\textwidth}
  \vspace{-0.18cm} 
  \hspace{-0.4cm}
  \includegraphics[width=1.0\linewidth]{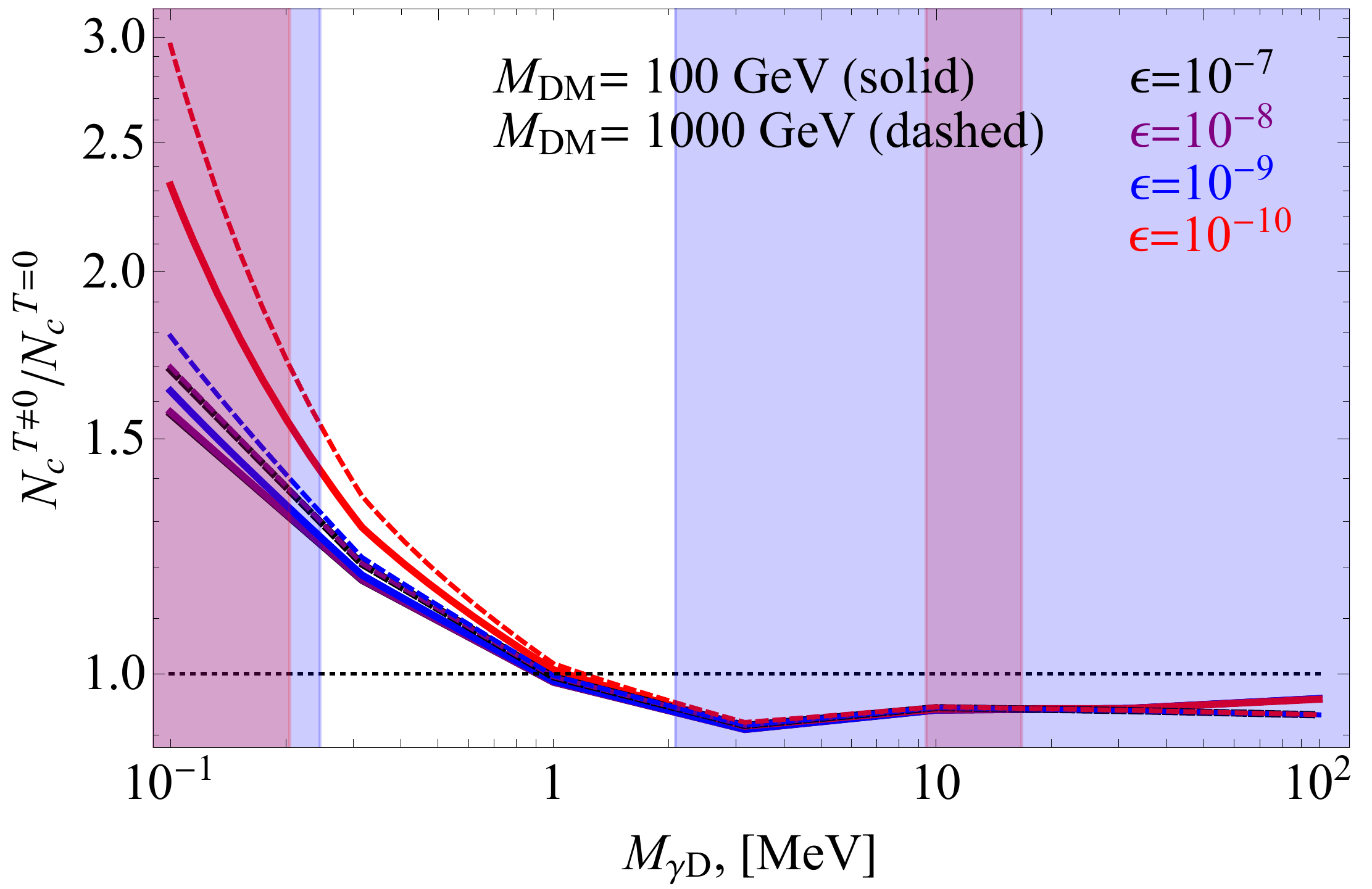}
\end{subfigure}
\caption{Ratio of the number of captured DM particles in the Sun at finite temperature to the zero-temperature result, $N^{T\neq 0}_{\text{c}}/N^{T=0}_{\text{c}}$, as a function of dark gauge boson mass $M_{\gamma_D}$. 
The solid (dashed) curves correspond to $M_{\text{DM}}= 100$ $(1000)$ GeV. The different colors correspond to different values of kinetic mixing parameter $\epsilon$. 
 Shaded blue (red) regions are ruled out by stellar cooling and supernova energy loss constraints for $\epsilon=10^{-9} (10^{-10})$; for $\epsilon=10^{-7}$ and $\epsilon=10^{-8}$ these constraints plus those from the experiments LSND and E137 rule out the entire mass range.}
\label{plot:SunNcThermalNoThermal}
\end{figure}

In Fig.~\ref{plot:SunNcThermalNoThermal} we show the ratio of the finite temperature calculation of the total bound population today $N_\text{c}\equiv N(\tau_{\text{Sun}})$ in the Sun to the zero-temperature calculation, as a function of $M_{\gamma_D}$.\footnote{Since both $C_{\text{c}}C_{\text{ann}}$ and $C_{\text{sc}}^2$ are proportional to the same power of $\alpha_D$, all dependence on $\alpha_D$ drops out of this population ratio.}  Here $\tau_{\text{Sun}} = 5\times 10^9$ years is the age of the Sun. 
For $M_{\gamma_D} \gg 10$ MeV, capture collisions are short-range and Fig.~\ref{plot:SunNcThermalNoThermal} accordingly displays the  expected small negative thermal corrections. 
Appreciable positive thermal corrections appear for $M_{\gamma_D}\lesssim $ MeV, more than an order of magnitude below $\Delta p_{typ}\sim 16$ MeV.
   For $\epsilon = 10^{-7}$, self-capture makes negligible contributions to the DM abundance, and the thermal enhancement exhibited for  $M_{\gamma_D} \lesssim$ MeV is due entirely to the enhanced nuclear capture rate.  As  $\epsilon$ is decreased, self-capture plays a more important role in determining the number of captured DM particles.  Since long-range self-capture also becomes more important at larger DM masses, we see the differences between the $M_{\text{DM}}=100$ GeV (solid) and $M_{\text{DM}}=1$ TeV (dashed) curves increase with decreasing $\epsilon$.  Thus the differences between the dashed and solid lines in Fig.~\ref{plot:SunNcThermalNoThermal}  at given values of $\epsilon$ and $M_{\gamma_D}$ help establish the relative importance of self-capture. We also discuss the regimes where nuclear capture or self-capture dominates in a broader range of parameter space in  Fig.~\ref{plot:Long_Range_Regulator} below.

For the values of $\epsilon$ shown in Fig.~\ref{plot:SunNcThermalNoThermal}, dark photons produced in the annihilations of captured DM escape the Sun before decaying \cite{Batell:2009zp, Schuster:2009au}.  Dark photons with masses above the MeV scale can decay to charged particle-antiparticle pairs outside the Sun, giving cosmic ray signals that can be probed by AMS \cite{Feng:2016ijc} and gamma-ray telescopes such as Fermi and HAWC \cite{Batell:2009zp,Ajello:2011dq,Leane:2017vag,Arina:2017sng,Albert:2018jwh,Niblaeus:2019gjk,Bell:2021pyy}.   In the parameter space relevant to AMS, $10^{-10}\lesssim \epsilon \lesssim 10^{-6}$ and $M_{\gamma_D}\in [1\text{ MeV}, 20\text{ MeV}]$, thermal corrections modify the $T=0$ approximation to the nuclear capture rate at the $\text{few }\%$ level. For these benchmark choices of $\alpha_D$ and $M_{\mathrm{DM}}$ and setting $M_{\gamma_D} = 100$ MeV, which corresponds to short-range nuclear capture, the values of the kinetic mixing parameter $\epsilon$ shown in Fig.~\ref{plot:SunNcThermalNoThermal} are at least an order of magnitude above the smallest coupling required for DM to thermalize with solar nuclei following the estimate of Ref.~\cite{Gaidau:2018yws}. For fixed $\epsilon$, going to lower mediator masses leads to an overall larger cross-section, but for sub-MeV mediators capture collisions are now in the transitional regime between long-range and short-range. Thus in this mediator mass range the first few collisions following capture may not be fully in the short-range regime. However, the subsequent nuclear scatterings in the approach to thermalization see successively smaller momentum transfers, and are thus well-described by short-range interactions.  The DM-nuclear cross-sections for the parameter points shown in Fig.~\ref{plot:SunNcThermalNoThermal} are well above the lower limits for short-range thermalization, and thus we expect the thermal DM population to be a self-consistent description.

In general, we expect from Fig.~\ref{plot:SunNcThermalNoThermal} that visibly-decaying mediators are not sufficiently light to realize large thermal enhancements to DM capture in the Sun. Sub-MeV mass mediators, which {\em can} give rise to sizeable thermal enhancements, are subject to stringent constraints.  In Fig.~\ref{plot:SunNcThermalNoThermal} we show the constraints from stellar cooling \cite{Redondo:2013lna,An:2013yfc,An:2014twa} and rapid energy loss in Supernova 1987A \cite{Chang:2016ntp} for $\epsilon = 10^{-9}$ and $\epsilon=10^{-10}$ (for  $\epsilon=10^{-7}$ and $10^{-8}$ the entire mass range is ruled out after imposing these constraints together with the exclusions from E137 \cite{Bjorken:1988as, Andreas:2012mt} and LSND \cite{Athanassopoulos:1997er,Essig:2010gu}).   For $\epsilon = 10^{-9}$ and $\epsilon=10^{-10}$ there are additional constraints on the minimal dark photon model from extragalactic background light \cite{Redondo:2008ec, Pospelov:2008jk} and supernovae \cite{DeRocco:2019njg} that can be circumvented in the presence of additional invisible decay modes for the dark photon.   In the regions of dark photon parameter space shown in Fig.~\ref{plot:SunNcThermalNoThermal}, stellar cooling and supernova energy loss constraints arise from the free-streaming of dark photons out of the star, and thus are not alleviated by invisible dark photon decay modes.

While for concreteness we have focused on dark photon mediators, stellar cooling constraints appear at similar mass scales for other possible mediators \cite{Hardy:2016kme,Knapen:2017xzo}, and thus these general conclusions apply more broadly.  
However, as Fig.~\ref{plot:SunNcThermalNoThermal} makes clear, sizeable thermal enhancements in solar capture begin to appear a relatively small distance below the exclusions from stellar cooling bounds.  Thus in hotter, more massive stars, characterized by larger values of $\Delta p_{typ}$, we can expect appreciable thermal enhancements to the capture of DM via MeV-scale mediator exchange.  Gravitational capture of DM in Population III stars is especially interesting in this regard.

 Population III stars are thought to form from the collapse of primordial gas clouds at the centers of DM minihaloes.  These first stars can have masses ranging up to $1000 M_\odot$, and are born in a DM-dense environment \cite{Bromm:2013iya,Klessen:2018fep}.  The rate of (short-range) DM gravitational capture in these massive early stars can potentially be large enough to give rise to detectable  impacts on the properties of the first-forming stars \cite{Freese:2015mta,Klessen:2018fep}; see also \cite{Ilie:2020nzp}.  
 Given the  large escape velocities within these massive stars, we expect thermal enhancements to long-range capture to be appreciable in these systems even for visibly-decaying mediators, and important for understanding their potential evolution.   Generally, the capture rate will be dominated by the central regions of the star where the escape velocity and temperature are both maximized; a model for the internal structure of Population III stars is therefore important for establishing the quantitative impacts of long-range capture in these objects.

\subsection{Solar capture through ultralight dark photons}
\label{sec:invis}

For dark photons, as distinct from other mediators, stellar cooling constraints become weaker as the dark photon mass decreases  \cite{An:2013yfc}.  Working in a basis of mass eigenstates, the suppression of dark photon emission with decreasing $M_{\gamma_D}$ from the plasma in the stellar core can be understood as an in-medium suppression of the effective mixing angle with the photon by the factor $M_{\gamma_D}^2/(M_{\gamma_D}^2-m_A^2)$, where $m_A$ is the effective mass for the photon (see, e.g., \cite{Knapen:2017xzo}).  The consequent suppression of the low-mass dark photon emission rate from stars opens up a distinct regime where gravitational capture can be mediated by very light (sub-eV) dark photons.  In this regime gravitational capture in the Sun does become long-range and can thus exhibit substantial thermal enhancements. 

\begin{figure}[t]
\begin{subfigure}{0.75\textwidth}
  \hspace{-.30cm}
  \includegraphics[width=1.0\linewidth]{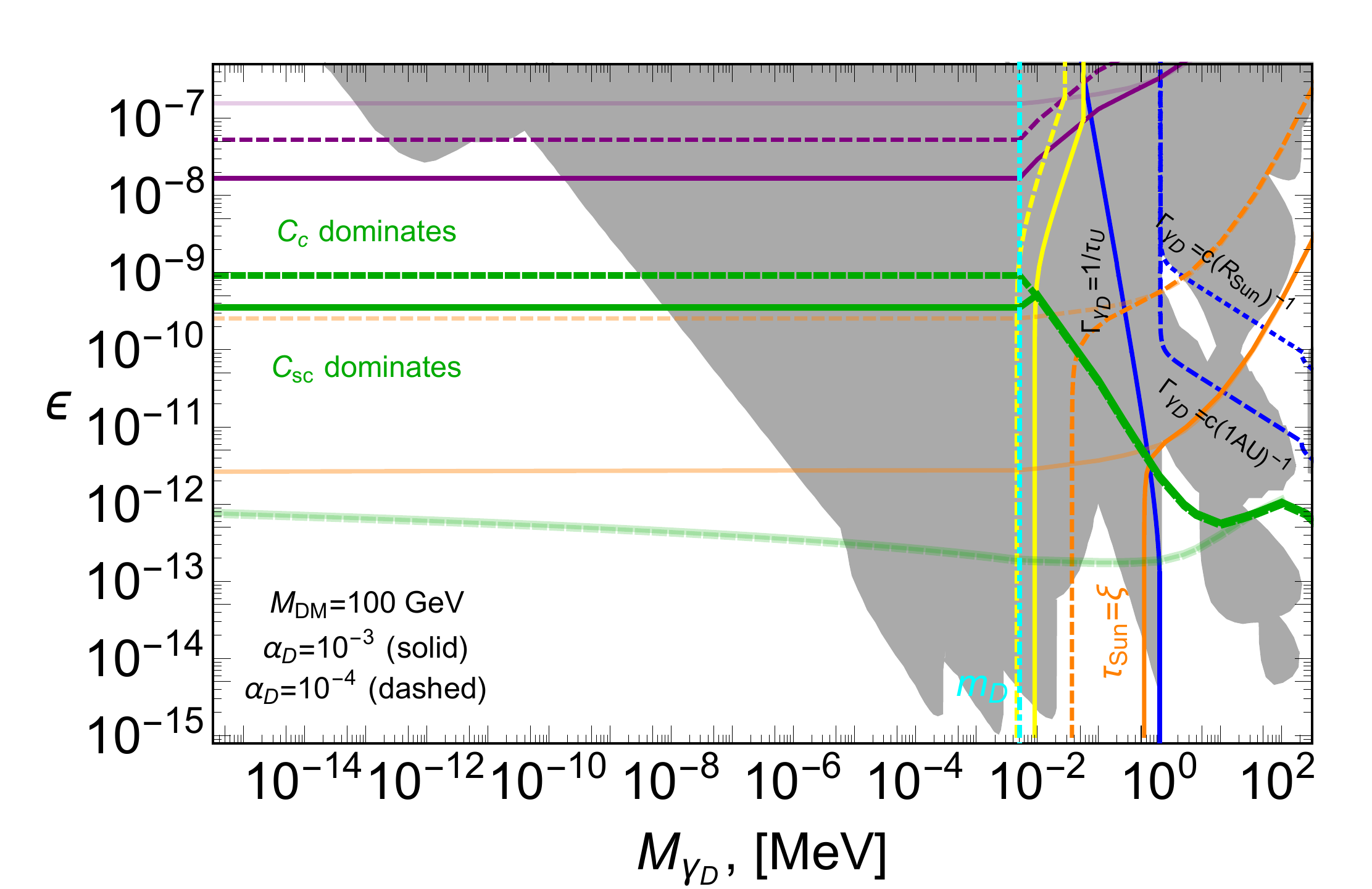}
\end{subfigure}%
\caption{Parameter space for light dark photons and gravitational capture of DM in the Sun.  The gray region shows the constraints on minimal dark photons from  stellar cooling \cite{Redondo:2013lna, An:2013yfc, An:2014twa}, Xenon10 \cite{An:2013yua}, supernova cooling \cite{Chang:2016ntp} and $e^+e^-$ pair production \cite{DeRocco:2019njg}, BBN \cite{Fradette:2014sza}, CMB spectral distortions \cite{Jaeckel:2008fi}, the intergalactic diffuse photon background \cite{Redondo:2008ec, Pospelov:2008jk}, the E137 \cite{Bjorken:1988as, Andreas:2012mt} and LSND \cite{Athanassopoulos:1997er, Essig:2010gu} experiments, and fifth force experiments \cite{Bartlett:1970js, Bartlett:1988yy}. The cyan line shows the solar Debye mass $m_D$, while blue lines indicate where the dark photon has a lifetime comparable to the age of the universe (solid),  the radius of Earth's orbit (dashed), and the Solar radius (dotted). Other contours indicate aspects of nuclear capture and self-capture of DM for the reference values  $M_{\text{DM}} = 100$ GeV and $\alpha_D = 10^{-3} \, (10^{-4})$ with solid (dashed) lines, as follows: bright purple lines indicate where the geometric limit for finite-temperature nuclear capture is saturated, while the pale purple contour indicates the geometric limit in the zero temperature calculation. Yellow lines indicate where the geometric limit for finite-temperature DM self-capture is saturated; the corresponding geometric limits in the zero-temperature calculation appear at values of $\epsilon$ above the shown parameter space. The bright (pale) orange contours indicate where the equilibration timescale $\xi$, Eq.~\ref{eq:xi}, is equal to the age of the Sun at finite (zero) temperature. Above the bright green line,  nuclear capture dominates the captured population dynamics and provides the dominant mechanism of gravitational capture, while below the line self-capture dominates. The pale green contour shows the analogous result in the zero-temperature calculation.
}
\label{plot:Long_Range_Regulator}
\end{figure}

For long-range scattering with $M_{\gamma_D}\gg m_A$, the parametric dependence of the capture coefficient $C_c$  on the model parameters $(M_{\text{DM}}, M_{\gamma_D}, \epsilon, \alpha_D)$ is given in Eq.~\ref{eq:ccscaling}.   We determine the coefficient of proportionality numerically. For dark photons with $M_{\gamma_D}\ll m_A$, we can incorporate the in-medium suppression of the SM coupling with the replacement $\epsilon^2/M_{\gamma_D}^2 \to \epsilon^2/m_A^2$. This scaling can be understood in the mass eigenstate basis by noting that within the solar plasma, the mostly-SM photon eigenstate picks up a coupling to the dark matter current given by $ \epsilon g_D m_A^2/(M_{\gamma_D}^2-m_A^2)$, which becomes $-\epsilon g_D$ in the limit $M_{\gamma_D}\ll m_A$.   Scattering through the mostly-SM eigenstate is then unsuppressed by the small ratio $(M_{\gamma_D}/m_A)$, and gives the  leading contribution to the capture rate in the light dark photon limit.

We estimate the size of the resulting IR cutoff by considering the leading contribution to long-range scattering in the nonrelativistic limit.  This contribution comes from the longitudinal portion of the photon propagator and in the nonrelativistic limit reduces to capture via an electric field. The Debye mass, related to the plasma frequency by $m_D^2 = \omega_p^2 (m/T) $, controls the long-range screening of electric fields, and has also been discussed as a regulating mass scale for long-range DM-SM scattering in \cite{Buen-Abad:2021mvc}.

We will assume, for simplicity, that all the nuclear species in the Sun are ionized and impose local charge neutrality. Then the plasma frequency is
\begin{align}
\omega_p^2 = \sum_i \frac{q_i^2 n_i}{m_i} \approx \frac{e^2 n_e}{m_e},
\label{OmegaPEqn} 
\end{align}
where the electron density in the Sun is
\begin{align}
n_e(r) = \sum_i Z_i \frac{\rho_i(r)}{m_i}.
\end{align}
Again, for simplicity we take the temperature and number densities at the core of the Sun, where most capture occurs.  Numerically this yields $\omega_p= 0.26$ keV and $m_D=5 $ keV. 
For $M_{\gamma_D} \gg \omega_p, m_D$ the long-range capture rate scales as $\epsilon^2/M_{\gamma_D}^2$; in the light dark photon parameter space, where $M_{\gamma_D} \ll \omega_p, m_D$, the capture rate becomes independent of the dark photon mass, scaling instead as $\epsilon^2/m_D^2$.

This scaling will hold up until reaching values of $\epsilon$ large enough to saturate the geometric limit. The geometric limit corresponds to the maximum physical capture rate where every DM particle incident on the Sun is captured.   We approximate this upper limit $C^{geom}_{\text{c}}$ as the incident flux of halo DM on the cross-sectional area of the Sun:
\begin{align}
\label{eq:geomlim}
C^{geom}_{\text{c}} = \pi R^2\frac{\rho_{\text{DM}}}{M_{\text{DM}}}\int d^3\vec{u}\frac{w(R)}{u}f_{\eta}(u)|\vec{w}(R)|.
\end{align} 
For fixed values of $M_{\text{DM}}$ and $\alpha_D$, setting
\begin{align}
C_{\text{c}}(M_{\text{DM}}, M_{\gamma_D}, \epsilon, \alpha_D)
= C^{geom}_{\text{c}}
\end{align}
then determines the curve in the $(M_{\gamma_D}, \epsilon)$ plane where the geometric limit is saturated. We show the resulting estimate for the saturation of the capture cross-section in Fig.~\ref{plot:Long_Range_Regulator} for a 100 GeV DM particle, with bright solid (dashed) purple lines indicating $\alpha_D=10^{-3}\,(10^{-4})$.  Above the bright purple line, the Sun is optically thick to DM and the capture cross-section has saturated; below the purple line, the capture rate is proportional to $(\epsilon/m_D)^{2}$.   For comparison, we show in pale purple the geometric saturation curve from the zero-temperature calculation for $\alpha_D=10^{-3}$.

Models featuring very light dark photons are restricted by constraints on the effective number of relativistic species during recombination, $\neff$.  These constraints can be ameliorated by simple models of late-time asymmetric reheating \cite{Chacko:2016hvu,Craig:2016lyx}, but the condition that the dark photon does not thermalize with the SM at late times requires the kinetic mixing to satisfy $\epsilon\lesssim 10^{-9}$ \cite{Vogel:2013raa,Chacko:2018vss}.   In this regime, the captured DM population will certainly thermalize with itself, but may or may not thermalize with the nuclei of the capturing body.  Thus the temperature of the  captured DM may in general differ from that of the nuclei, but we nonetheless expect a thermal distribution for both populations to remain an excellent description. 

For simplicity we will consider the interplay between capture and self-capture assuming that the captured DM has thermalized with the nuclei at $T_{\text{Sun}}$. We expect this to be a conservative assumption in the sense that when a captured DM particle is unable to efficiently transfer momentum to nuclei, the captured DM ball will have a {\em higher} temperature than the SM, and thus self-capture will become even more important relative to nuclear capture.

Let us first consider the geometric limit of self-capture.  Letting $r_x$ denote the radius of the captured DM ball, adapting Eq.~\ref{eq:geomlim} to self-capture collisions gives
\begin{align}
C_{\text{sc}}^{geom} = \pi r_x^2 \frac{\rho_{\text{DM}}}{M_{\text{DM}}}\int d^3\vec{u}\frac{w(r_x)}{u}f_{\eta}(u)|\vec{w}(r_x)| .
\end{align}
We define $r_x$  as the radius of a sphere enclosing 95\% of the captured DM particles:
\begin{align}
0.95 = \int_0^{r_x} dV n_c(r).
\end{align}
For $M_{\text{DM}}=100\text{ GeV}$, $r_x = 0.02R$. The total self-capture rate $C_{\text{sc}}N_c$ is bounded from above by the geometric self-capture rate:
\begin{align}
C_{\text{sc}}(M_{\text{DM}}, M_{\gamma_D}, \alpha_D)N_c(M_{\text{DM}}, M_{\gamma_D}, \alpha_D, \epsilon)\leq C_{\text{sc}}^{geom},
\end{align}
where the right-hand side only depends on $M_{\text{DM}}$ and $N_c$ is given by Eq.~\ref{eq:DMCapturedPopulation}. In Fig.~\ref{plot:Long_Range_Regulator} yellow solid (dashed) lines show the curves in the $(M_{\gamma_D}, \epsilon)$ plane that saturate this finite-temperature geometric limit  for fixed $M_{\text{DM}}=100$ GeV and $\alpha_D=10^{-3} \,(10^{-4})$.  In the parameter space of interest, the geometric limit is saturated at values of the dark photon masses that are much larger than regulators from in-medium effects.

 Also in Fig.~\ref{plot:Long_Range_Regulator},  we show in orange where the equilibration time scale $\xi$ is equal to $\tau_{\text{Sun}}$; above the orange curves, the DM population has reached a steady-state value.  Green lines indicate where $C_{\text{c}}C_{\text{ann}}= C_{\text{sc}}^2/4$; above (below) the green lines, $C_{\text{c}}C_{\text{ann}}\gg C_{\text{sc}}^2/4$ ($C_{\text{c}}C_{\text{ann}}\ll C_{\text{sc}}^2/4$) and nuclear capture (self-capture) is the dominant mechanism for gravitational DM capture.

The finite-temperature calculation shows three important differences from the zero-temperature calculation.  First, vast regions of parameter space are in the geometric limit for self-capture, a more drastic change than exhibited for the case of nuclear capture. An immediate consequence is that the much larger self-capture rate is sufficient to equilibrate the captured population at much smaller values of $\epsilon$ than the zero-temperature calculation would suggest. Finally,  owing to the different parametric scalings of capture and self-capture at finite temperature, self-capture dominates over nuclear capture in an expanded range of parameter space.

Ultra-light dark photons are cosmologically long-lived.  In this light dark photon regime, DM annihilation within massive bodies therefore produces no visible signals from mediator decays; however, the potentially very large captured DM population may leave an observable imprint in stellar structure or energy transport.  Concrete realizations of this scenario have been considered  in the context of mirror stars and give rise to astrophysically interesting signatures and constraints already in the zero-temperature regime \cite{Curtin:2019ngc}; we expect the thermal enhancements we find here will make such probes of mirror dark matter even more powerful.

\section{Conclusions}
\label{s.concl}

We have reexamined the gravitational capture of DM by non-degenerate astrophysical bodies at finite temperature. When capture occurs via long-range interactions, we find that retaining the thermal motion of the target particles leads to a capture rate that is quadratically dependent on the mass of the mediator.  This result gives a potentially sizable enhancement to the DM capture rate compared to expectations from the zero-temperature calculation, which predicts a logarithmic sensitivity to the mediator mass.  This parametric difference between the finite-temperature and zero-temperature results can be simply understood by observing that the small volume of phase space that results in capture via arbitrarily soft scattering at finite target temperature shrinks to a single point in the zero-temperature limit.  As a concrete example, we provide numerical calculations for the solar capture of a Dirac DM particle interacting with the SM via a kinetically mixed dark photon, and study the transition from short-range to long-range capture regimes.   

Numerically we find that substantial thermal enhancements to the DM capture rate begin to appear once the mediator mass is more than an order of magnitude smaller than the typical momentum transfer in DM-nucleus collisions.  Since we estimate the typical momentum transfer scale to be $\Delta p_{typ}\sim$ 16~MeV in the Sun, sizeable thermal enhancements to solar capture are not compatible with visible decays of the mediator.  However, for light dark photons, substantial enhancements can be realized, in nuclear capture but particularly in self-capture.   We study the interplay between nuclear capture and self-capture in the Sun and demonstrate that in almost all of  the relevant parameter space, the captured DM population becomes optically thick to self-capture.   Thus we expect stellar capture of mirror dark matter to be substantially more efficient than previously understood. Even for MeV-scale mediators, we expect the thermal enhancement to long-range capture to be important for  understanding DM capture in the first generation of stars, which are substantially more massive than the Sun.

\subsection*{Acknowledgements}
We thank David Curtin for useful conversations. The work of CG and JS is supported in part by DOE
Early Career grant DE-SC0017840.


\appendix

\section{The angular integral $J_c$}
\label{appendix.A}
In this appendix we present the details of the analytical and numerical calculation of the angular integral $J_c$, defined in Eq.~\ref{Eqn:Jc}:
\begin{align}
J_c\equiv\frac{1}{4\pi}\int_0^{2\pi}d\phi_{CM}\int_{-1}^{1} d\cos\theta \int_0^{2\pi} d\phi_{\chi}\frac{1}{(1-\cos\theta +\mu)^2}\Theta (\beta(r) - \cos\gamma),
\end{align}
with $\cos\gamma = \sin\alpha\sin\theta\cos (\phi_{\chi}-\phi_{CM}) + \cos\alpha\cos\theta$. Note that $\cos\gamma$ depends on azimuthal angles only through the difference $\phi_{\chi}-\phi_{CM}$. Switching variables to $\phi_{\pm} = \phi_{CM}\pm\phi_{\chi}$, we write
\begin{align}
J_c &= \frac{1}{4\pi}\int_{-1}^{1}d\cos\theta\frac{1}{(1-\cos\theta +\mu)^2}\int d\phi_+ \int d\phi_- \frac{1}{2} \Theta\left(2\pi-\frac{1}{2}(\phi_+ +\phi_-)\right) \times\nonumber\\
&\times \Theta\left(2\pi-\frac{1}{2}(\phi_+ -\phi_-)\right)\Theta\left(\frac{1}{2}(\phi_+ +\phi_-)\right) \Theta\left(\frac{1}{2}(\phi_+ -\phi_-)\right)\Theta(\beta(r) -\cos\gamma),
\label{eqn:AppCAux1}
\end{align}
where the new factor of 1/2 comes from the Jacobian of the transformation and the first four step functions impose the original bounds of integration $\phi_{CM}, \phi_{\chi} \in [0, 2\pi )$. The latter two of the aforementioned step functions demand that $\phi_+>0$. 
The product of the first four step functions in Eq.~\ref{eqn:AppCAux1} can be turned into upper and lower limits for the $\phi_\pm$ integration,
\begin{align}
\int_0^{\infty}d\phi_+\int_0^{\infty}&d\phi_-\Theta\left(4\pi-(\phi_+ +\phi_-)\right)\Theta\left(4\pi-(\phi_+ -\phi_-)\right)\Theta\left(\phi_+ +\phi_-\right) \Theta\left(\phi_+ -\phi_-\right)= \nonumber \\
&=\int_0^{\infty}d\phi_-\int_{\phi_-}^{4\pi -\phi_-}d\phi_+ \Theta(4\pi -2\phi_-) =\int_0^{2\pi}d\phi_-(4\pi -2\phi_-).
\label{eqn:AppCAux3}
\end{align}
We further exploit the symmetry of this integral by partitioning the integration region into three pieces, $0\leq \phi_- < \pi/2$, $\pi/2\leq \phi_- < 3\pi/2$, and $3\pi/2\leq \phi_-<2\pi$.
In the last region we shift $\phi_-\rightarrow \phi_- -2\pi$, under which $\cos\phi_-\rightarrow \cos\phi_-$ and $\cos\gamma\rightarrow \cos\gamma$. In the second region, consider $\phi_-\rightarrow \phi_- -\pi$, which implies $\cos\phi_-\rightarrow -\cos\phi_-$ and $\cos\gamma\rightarrow \cos\bar{\gamma} \equiv -\sin\alpha \sin\theta \cos\phi_- +\cos\alpha \cos\theta$. Then,
\begin{align}
\int_0^{2\pi}d\phi_-(4\pi -2\phi_-) & \Theta(\beta(r)-\cos\gamma)= \int_0^{\pi/2}d\phi_-(4\pi)\Theta(\beta(r)-\cos\gamma) + \nonumber\\
&+ \int_0^{\pi/2}d\phi_-(-2\phi_-)\Theta(\beta(r)-\cos\gamma) +  \int_{-\pi/2}^{\pi/2}d\phi_-(2\pi)\Theta(\beta(r)-\cos\bar{\gamma})+ \nonumber \\
&+\int_{-\pi/2}^{\pi/2}d\phi_-(-2\phi_-)\Theta(\beta(r)-\cos\bar{\gamma}) + \int_{-\pi/2}^0d\phi_-(-2\phi_-)\Theta(\beta(r)-\cos\gamma).
\label{egn:7.6}
\end{align}
The fourth integral vanishes by symmetry. The sum of the second and fifth integrals vanishes for the same reason. The third integral is symmetric around zero, so that we can simplify $J_c$ to the following expression:
\begin{align}
J_c = \int_0^{\pi/2}d\phi_-\int_{-1}^{1}d\cos\theta\frac{1}{(1-\cos\theta +\mu)^2}\left[\Theta(\beta(r)-\cos\gamma)+\Theta(\beta(r)-\cos\bar{\gamma})\right].
\end{align}
We note that $J_c = J_c(\beta(r), \cos\alpha, \mu)$, with $-1\leq\beta(r)\leq 1$ and $\beta(r) \leq \cos\alpha \leq 1$. We can take advantage of this property to reduce the dimension of the numerical integration of $C_{\text{c}}$. We precompute $J_c$ on the following discretized grid: $\beta \in [-1, 1]$ in 60 steps, $\cos\alpha \in [-1, 1]$ in 70 steps and $\mu \in [10^{-15.5}, 10^{12.5}]$ in 28 log steps. The additional step function $\Theta(\cos\alpha-\beta(r))$ in  Eq.~\ref{Eqn:dCdVAux1} is included in the subsequent numerical computation over the particle speeds. Loosely speaking, this procedure reduces a 6-dimensional integral to a 4$\oplus$2-dimensional integral, significantly reducing the computational cost.

\section{The angular integral $J_{sc}$} 
\label{appendix.B}

In this appendix we discuss the details of the analytic and numerical calculation of the angular integral introduced in the calculation of the self-capture coefficient. We begin with the calculation of $J_{sc}$, which was defined in Eq.~\ref{defn:Jsc},
\begin{align}
J_{sc} = \frac{1}{4\pi}\int_0^{2\pi}d\phi_{CM}\int_0^{2\pi}d\phi_{\chi}\int_{-1}^{1}d\cos\theta f_{sc}(\cos\theta, \mu )\Theta (\beta(r) -|\cos\gamma |),
\end{align}
with Eq.~\ref{defn:fsc} defining $f_{sc}$. We note again that only the difference of the azimuthal angles appears in the integrand through $\cos\gamma$. Therefore, we can apply the same argument used in Appendix~\ref{appendix.A}, by switching to $\phi_{\pm} = \phi_{CM}\pm\phi_{\chi}$:
\begin{align}
J_{sc} = \frac{1}{4\pi}\int_0^{2\pi}d\phi_{-}(4\pi -2\phi_-)\int_{-1}^{1}d\cos\theta f_{sc}(\cos\theta, \mu )\Theta (\beta(r) -|\cos\gamma |).
\end{align}
We continue by applying the same logic as in Eq.~\ref{egn:7.6}, where we partition and shift the $\phi_-$ integration region to obtain
\begin{align}
J_{sc} = \int_0^{\pi/2}d\phi_{-}\int_{-1}^{1}d\cos\theta f_{sc}(\cos\theta, \mu )\left(\Theta (\beta(r) -|\cos\gamma |)+\Theta (\beta(r) -|\cos\bar{\gamma}|)\right),
\end{align}
where we recall that $\cos\bar{\gamma}\equiv -\sin\alpha \sin\theta \cos\phi_- +\cos\alpha \cos\theta$. Next we define $f_{sc}^{sym}(\cos\theta , \mu) =  f_{sc}(\cos\theta , \mu)+  f_{sc}(-\cos\theta , \mu)$ and reduce the domain of integration of $\cos\theta$ by using the symmetry properties of the integrand
\begin{align}
J_{sc} = \int_0^{\pi/2}d\phi_{-}\int_{0}^{1}d\cos\theta f_{sc}^{sym}(\cos\theta, \mu )\left(\Theta (\beta(r) -|\cos\gamma |)+\Theta (\beta(r) -|\cos\bar{\gamma}|)\right).
\end{align}
Note that $J_{sc}=J_{sc}(\beta(r), \cos\alpha, \mu)$, with $0\leq\beta(r)\leq \cos\alpha \leq 1$ (see Eq.~\ref{eqn:Cc313}). Similar to $J_c$, we calculate $J_{sc}$ by discretizing it on the following grid: $\beta \in [0, 1]$ in 60 steps, $\cos\alpha \in [0, 1]$ in 70 steps and $\mu \in [10^{-15.5}, 10^{12.5}]$ in 28 log steps.

\section{Dark Matter self-evaporation}
\label{appendix.C}

In this appendix we discuss the calculation of the self-evaporation rate. Self-evaporation occurs when two captured DM particles scatter such that one of the particles becomes gravitationally unbound after the collision. We stress that self-evaporation does not occur in the zero-temperature approximation, since the captured DM particles have no kinetic energy to overcome the potential well. Therefore, one is confronted with incorporating thermal motion of the target particles from the outset. 

\par
The calculation proceeds in analogy with Secs.~\ref{s.nuclear_cap} and~\ref{s.selfcaptAndselfejec}. We will compute the total self-evaporation rate, accounting for particle ($\chi$) and antiparticle ($\bar{\chi}$) self-evaporation by scattering against both particles and antiparticles,
\begin{align}
C_{\text{sevap}}= \int_0^R dr 4\pi r^2 \left(\frac{dC^{\chi\chi}_{\text{sevap}}}{dV}+\frac{dC^{\chi\bar{\chi}}_{\text{sevap}}}{dV}+\frac{dC^{\bar{\chi}\chi}_{\text{sevap}}}{dV}+\frac{dC^{\bar{\chi}\bar{\chi}}_{\text{sevap}}}{dV}\right), 
\end{align}
where
\begin{align}
\frac{dC^{\chi\bar{\chi}}_{\text{sevap}}}{dV} = n^{\chi}_c(r)n^{\bar{\chi}}_c(r)\int_0^{v_{\text{esc}}(r)}d^3\vec{v}_1f_c(\vec{v}_1)\int_0^{v_{\text{esc}}(r)}d^3\vec{v}_2f_c(\vec{v}_2)|\vec{v}_1-\vec{v}_2|\sigma^{\chi\bar{\chi}}_{\text{sevap}}
\end{align}
is the self-evaporation per unit volume of particles by scattering against antiparticles, and the other three $dC_{\text{sevap}}/dV$ terms are defined analogously. The colliding DM particles, with initial velocities $\vec{v}_1$ and $\vec{v}_2$, are described by the same Maxwell distribution at the core temperature of the capturing body, given by Eq.~\ref{eqn:DMdistroMB}. In the case where the particle and antiparticle populations are equal, $n^{\chi}_c(r) = n^{\bar{\chi}}_c(r)=1/2 \, n_c(r)$, where $n_c(r)$ is the total number density of captured DM (see Eq.~\ref{eqn:Csc1}), and we write
\begin{align}
&C_{\text{sevap}}= \frac{1}{4}\int_0^R dr 4\pi r^2 \frac{dC_{\text{sevap}}}{dV}, \nonumber \\
\frac{dC_{\text{sevap}}}{dV} = n^2_c(r) &\int_0^{v_{\text{esc}}(r)}d^3\vec{v}_1f_c(\vec{v}_1)\int_0^{v_{\text{esc}}(r)}d^3\vec{v}_2f_c(\vec{v}_2)|\vec{v}_1-\vec{v}_2|\sigma_{\text{sevap}}.
\end{align}
We begin by transforming the kinematic integrals to the CM frame, where we use the definitions introduced in Eq.~\ref{CM:vel},
\begin{align}
 \int_0^{v_{\text{esc}}(r)}&d^3\vec{v}_1f_c(\vec{v}_1)\int_0^{v_{\text{esc}}(r)}d^3\vec{v}_2f_c(\vec{v}_2) = \nonumber \\
 =\int_0^{\infty} & d^3\vec{v}_1 \Theta ( v_{\text{esc}}(r)-|\vec{v}_1|)f_c(\vec{v}_1)\int_0^{\infty}d^3\vec{v}_2\Theta ( v_{\text{esc}}(r)-|\vec{v}_2|)f_c(\vec{v}_2) = \nonumber \\
= \int_0^{\infty}& d^3\vec{v}_{CM}\int_0^{\infty}d^3\vec{v}'_1 J f_c(\vec{v}_{CM}+\vec{v}'_1)f_c(\vec{v}_{CM}-\vec{v}'_1)\times \nonumber\\
&\times\Theta ( v_{\text{esc}}(r)-|\vec{v}_{CM}+\vec{v}'_1|)\Theta ( v_{\text{esc}}(r)-|\vec{v}_{CM}-\vec{v}'_1|),
\label{eqn:JsevapAux0}
\end{align}
where $J=8$ is the Jacobian of the transformation. First, we note the following simplification for the exponentials appearing in the product of the two Maxwell distributions:
\begin{align}
e^{-\frac{M_{\text{DM}}}{2T_{\text{Sun}}}(v^2_{CM} +v^{'2}_1+2\vec{v}_{CM}\cdot\vec{v}'_1)} e^{-\frac{M_{\text{DM}}}{2T_{\text{Sun}}}(v^2_{CM} +v^{'2}_1-2\vec{v}_{CM}\cdot\vec{v}'_1)} = e^{-\frac{M_{\text{DM}}}{2T_{\text{Sun}}}(2v^2_{CM} +2v^{'2}_1)} .
\end{align}
As before, we orient the $z$-axis of our frame along $\vec{v}'_1$ such that the solid angle integration of $d^3\vec{v}'_1$ proceeds trivially. We let $\alpha$ and $\phi_{CM}$ specify the polar and azimuthal angles of $\vec{v}_{CM}$ with respect to the $z$-axis. With this convention, the two step functions $\Theta ( v_{\text{esc}}(r)-|\vec{v}_{CM} \pm\vec{v}'_1|)$ in the kinematic integrals of Eq.~\ref{eqn:JsevapAux0} turn into bounds on $c\alpha\equiv \cos\alpha$: 
\begin{align}
 v^2_{\text{esc}}(r) \geq v^2_{CM}+v^{'2}_1 \pm 2v_{CM}v'_1c\alpha,
\end{align}
which we write compactly as
\begin{align}
|c\alpha|\leq \beta (r),
\end{align}
with $\beta$ defined in Eq.~\ref{defn:beta}. Collecting these calculations together,
\begin{align}
\frac{dC_{\text{sevap}}}{dV} = n^2_c(r)\int_0^{\infty} & v^2_{CM}dv_{CM}\int_0^{\infty}v^{'2}_1dv'_1\int_{-1}^{1}dc\alpha\Theta (\beta (r) -|c\alpha|) \times \nonumber \\
& \times \int_0^{2\pi}d\phi_{CM}J4\pi f_c(\sqrt{2}v_{CM})f_c(\sqrt{2}v'_1) 2v'_1\sigma_{\text{sevap}}.
\label{eqn:JsevapAux1}
\end{align}
Next we calculate the self-evaporation cross-section in the non-relativistic limit.  For $\chi\chi$ and $\bar{\chi}\bar{\chi}$ scattering, both $t$- and $u$-channel diagrams contribute.  The rate for these processes includes a symmetry factor of $1/2$ to account for the identical initial state particles, which for simplicity we incorporate into an effective matrix element.  For $\bar{\chi}\chi$ and $\chi\bar{\chi}$ self-evaporation only the $t$-channel diagram contributes.
Combining all four processes, we can write the effective matrix element-squared as 
\begin{align}
&\quad\quad\quad |\mathcal{M}|^2 = \frac{4g_D^4}{v^{'4}_1}f_{\text{sevap}}(\cos\theta, \mu), \text{ where} \\
f_{\text{sevap}}(\cos\theta, \mu) = & \frac{3}{(1-\cos\theta +\mu)^2} +\frac{1}{(1+\cos\theta +\mu)^2} +\frac{1}{(1-\cos\theta+\mu)(1+\cos\theta+\mu)}.
\end{align}
Self-evaporation occurs when one of the particles has a final speed larger than the local escape speed. This kinematic constraint is 
\begin{align}
\Theta_{\text{sevap}} =\Theta (|\vec{v}^{lab}_{1,f}|-v_{\text{esc}}(r)) =\Theta(\cos\gamma- \beta (r)),
\end{align}
where the last equality is written in terms of  the angle $\gamma$, defined in Eq.~\ref{defn:gamma}. We convolute this step function with the matrix element above and integrate over the solid angle of the final state particles to obtain the self-evaporation cross-section,
\begin{align}
\sigma_{\text{sevap}}= \frac{1}{256\pi^2M^2_{\text{DM}}} &\int d\cos\theta d\phi_{\chi}|\mathcal{M}|^2\Theta_{\text{sevap}} = \nonumber \\
&= \frac{\alpha^2_D}{4M^2_{\text{DM}}v^{'4}_1} \int d\cos\theta d\phi_{\chi}f_{\text{sevap}}(\cos\theta, \mu)\Theta(\cos\gamma- \beta (r)).
\label{eqn:JsevapAux2}
\end{align}
Inserting Eq.~\ref{eqn:JsevapAux2} in Eq.~\ref{eqn:JsevapAux1} and collecting terms gives the following result for self-evaporation per unit volume:
\begin{align}
\frac{dC_{\text{sevap}}}{dV} = n^2_c(r)\int_0^{\infty}v^2_{CM}dv_{CM}& \int_0^{\infty}v^{'2}_1dv'_1 \Theta (\beta (r)) \Theta (1-\beta (r))\times \nonumber \\
&\times (4\pi)^2 J f_c(\sqrt{2}v_{CM})f_c(\sqrt{2}v'_1) 2v'_1\frac{\alpha^2_D}{4M^2_{\text{DM}}v^{'4}_1} J_{\text{sevap}}(\beta (r), \mu),
\label{eqn:JsevapAux3}
\end{align}
where we introduced the angular integral
\begin{align}
J_{\text{sevap}}(\beta (r), \mu) \equiv \frac{1}{4\pi}\int_0^{2\pi}d\phi_{CM} &\int_0^{2\pi}d\phi_{\chi}\int_{-1}^1dc\alpha \int_{-1}^1d\cos\theta \times \nonumber \\
&\times f_{\text{sevap}}(\cos\theta, \mu)\Theta (\beta (r) -|c\alpha|)\Theta(\cos\gamma- \beta (r)).
\label{Jsevap}
\end{align}
The two step functions in Eq.~\ref{Jsevap} impose $0\leq\beta(r)\leq 1$ for non-zero $J_{\text{sevap}}$. For clarity, we further explicitly include this condition in Eq.~\ref{eqn:JsevapAux3}. In what follows we consider the computation of $J_{\text{sevap}}$. We apply the transformation  $\phi_{\pm} = \phi_{CM}\pm\phi_{\chi}$, outlined in Eqs.~\ref{eqn:AppCAux1}--\ref{eqn:AppCAux3}, to obtain
\begin{align}
J_{\text{sevap}}(\beta (r), \mu) = \frac{1}{4\pi}\int_{-1}^1dc\alpha &\int_0^{2\pi}d\phi_-(4\pi-2\phi_-)\int_{-1}^1d\cos\theta \times \nonumber \\
&\times\Theta (\beta (r) -|c\alpha|)f_{\text{sevap}}(\cos\theta, \mu)\Theta(\cos\gamma- \beta (r)),
\end{align}
followed by the symmetry argument used in Eq.~\ref{egn:7.6},
\begin{align}
J_{\text{sevap}}(\beta (r), \mu) = \int_0^{\pi/2}& d\phi_-\int_{-1}^1dc\alpha \int_{-1}^1d\cos\theta \Theta (\beta (r) -|c\alpha|) \times \nonumber \\
&\times f_{\text{sevap}}(\cos\theta, \mu)\big(\Theta(\cos\gamma- \beta (r))+\Theta(\cos\bar{\gamma}- \beta (r))\big),
\label{eqn:AppCAux9}
\end{align}
where we recall that $\cos\bar{\gamma}\equiv -s\alpha \sin\theta \cos\phi_- +c\alpha \cos\theta$. We note, however, that $\Theta(\cos\bar{\gamma}- \beta (r))$ is always zero in the given volume of integration. To see this, note that $\Theta (\beta (r) -|c\alpha|)$ imposes $c\alpha \cos\theta\leq\beta (r)$, where we also recall the kinematic constraint $0\leq\beta\leq 1$. Therefore, $\cos\bar{\gamma}=-s\alpha \sin\theta \cos\phi_- +c\alpha \cos\theta \leq \beta$ since the first term is non-positive in the integration region. We write the $\Theta(\cos\gamma- \beta (r))$ step function as a quadratic inequality for $c\alpha$:
\begin{align}
0> ( &\cos^2\theta + \sin^2\theta \cos^2\phi_-)c^2\alpha -2\beta (r) \cos\theta c\alpha +\beta^2(r) -\sin^2\theta \cos^2\phi_-,
\end{align}
which is satisfied for $c\alpha\in(c\alpha_-, c\alpha_+)$, where 
\begin{align}
c\alpha_{\pm} = \frac{\beta (r)\cos\theta \pm \sqrt{\sin^2\theta \cos^2\phi_-(\cos^2\theta +\sin^2\theta \cos^2\phi_- -\beta^2(r))}}{\cos^2\theta +\sin^2\theta \cos^2\phi_-}.
\end{align}

\begin{figure}[t]
\begin{subfigure}{0.5\textwidth}
  \hspace{-.30cm}
  \includegraphics[width=1.0\linewidth]{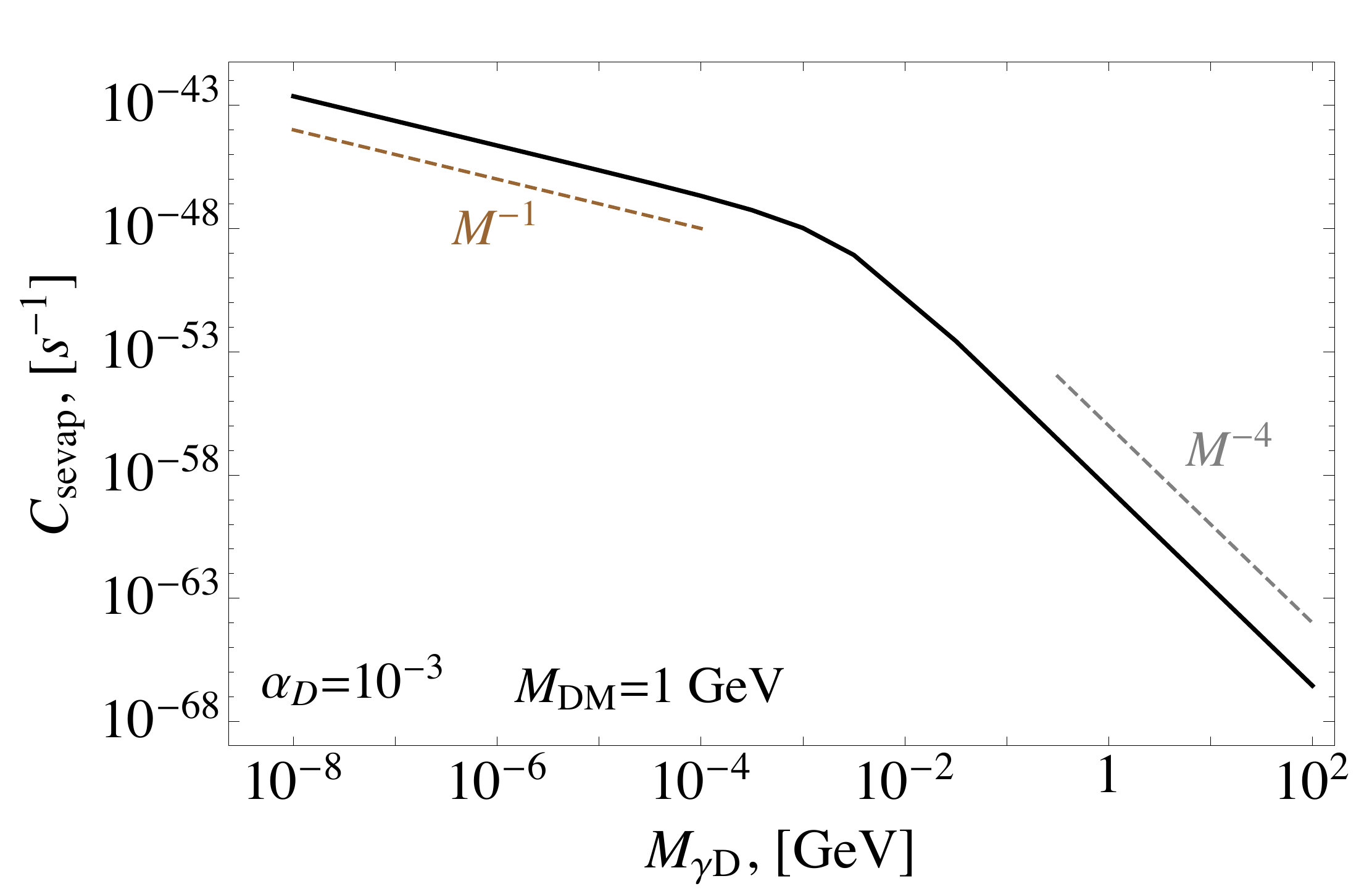}
\end{subfigure}%
\begin{subfigure}{0.480\textwidth}
  \hspace{.0cm}
  \vspace{-0.22cm}
  \includegraphics[width=1.0\linewidth]{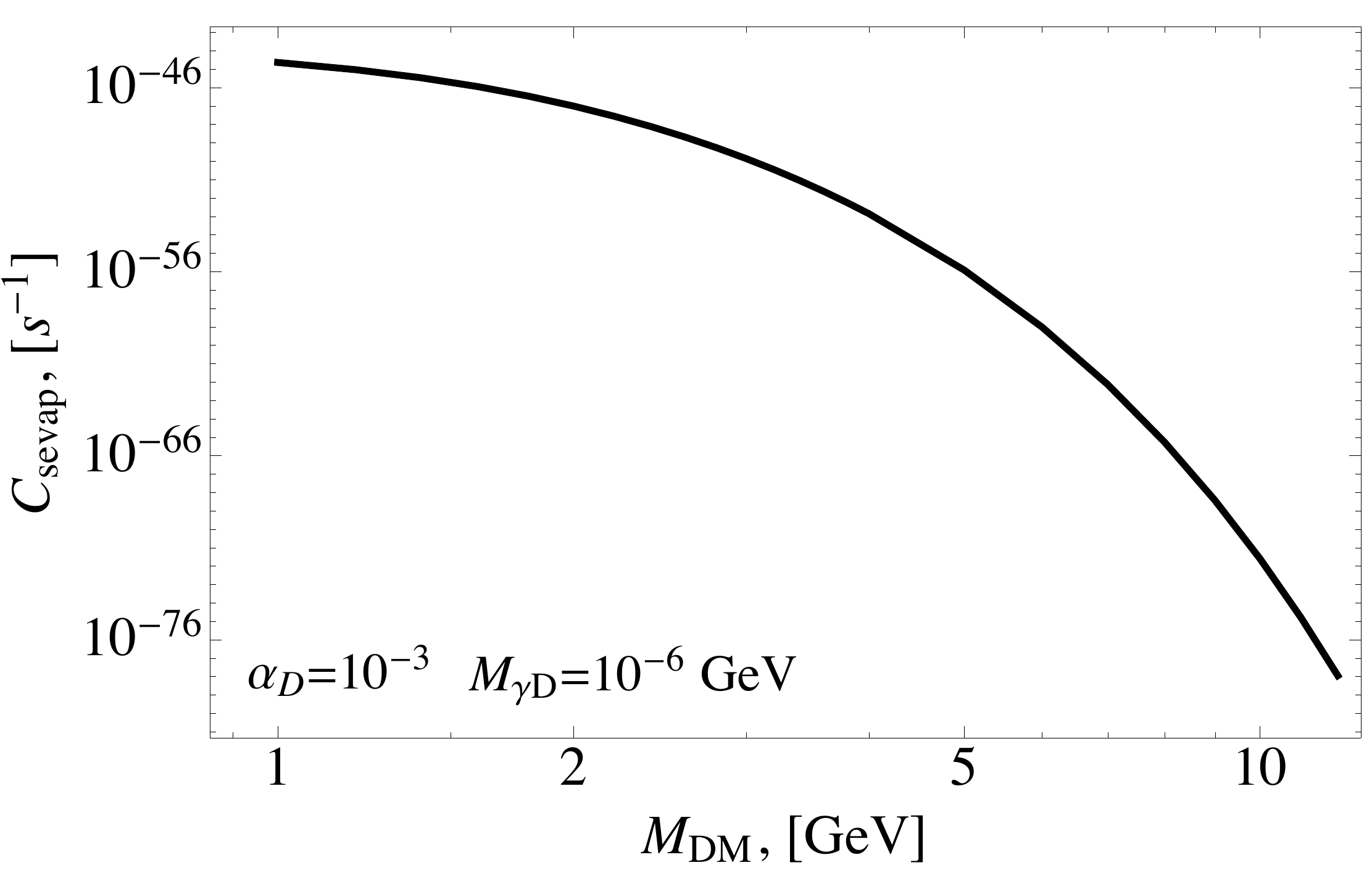}
\end{subfigure}
\caption{{\bf Left:} Self-evaporation coefficient in the Sun as a function of mediator mass $M_{\gamma_D}$. Dashed lines are included to highlight the parametric scaling with $M_{\gamma_D}$ in the long- and short-range regimes. {\bf Right:} Self-evaporation coefficient in the Sun as a function of DM mass.  We fix $M_{\gamma_D}=10^{-6}$ GeV, corresponding to long-range scattering, and $\alpha_D=10^{-3}$.}
\label{plot:SunEarthCsevap}
\end{figure}

\noindent
To ensure that $c\alpha_{\pm}$ are real we impose the positivity of the discriminant with $\Theta (\cos^2\theta +\sin^2\theta \cos^2\phi_- -\beta^2(r))$. The $c\alpha$ integral can now be evaluated in closed form, producing four terms which correspond to different orderings of upper and lower limits of integration on  $c\alpha$  imposed by the $\Theta (\beta (r) -|c\alpha|)$ and $\Theta(c\alpha_+-c\alpha)\Theta(c\alpha - c\alpha_-)$ step functions.  Inserting this calculation in Eq.~\ref{eqn:AppCAux9} gives
\begin{align}
J_{\text{sevap}}&(\beta (r), \mu) = \int_0^{\pi/2} d\phi_-\int_{-1}^1d\cos\theta f_{\text{sevap}}(\cos\theta, \mu)\Theta (\cos^2\theta +\sin^2\theta \cos^2\phi_- -\beta^2(r))\times \nonumber \\
\times \big[& (c\alpha_+-c\alpha_-)\Theta(\beta (r)-c\alpha_+)\Theta(c\alpha_- +\beta (r)) + 2\beta (r) \Theta(c\alpha_+-\beta (r))\Theta(-\beta (r)-c\alpha_-) \nonumber + \\
& + (c\alpha_++\beta(r))\Theta(\beta(r)-c\alpha_+)\Theta(-\beta(r)-c\alpha_-)\Theta(c\alpha_+ +\beta(r))  + \nonumber \\
& + (\beta(r)-c\alpha_-)\Theta(c\alpha_+-\beta(r))\Theta(c\alpha_-+\beta(r))\Theta(\beta(r)-c\alpha_-)   \big].
\end{align}
We take advantage of the restricted functional dependence $J_{\text{sevap}}(M_{\text{DM}}, M_{\gamma_D}, r, v_{CM},v'_1) = J_{\text{sevap}}(\beta(r), \mu)$, with $\beta(r)$ in the finite range $0\leq\beta(r)\leq1$, to precompute $J_{\text{sevap}}$ numerically by placing it on the following grid: $\beta \in [0, 1]$ in 70 even steps and $\mu \in [10^{-15.5}, 10^{12.5}]$ in 28 log steps. We collect all calculations to write the final expression for the self-evaporation coefficient
\begin{align}
C_{\text{sevap}} = (4\pi)^3\frac{\alpha^2_D}{M^2_{\text{DM}}}& \int_0^R dr r^2 n^2_c(r)\int_0^{v_{\text{esc}}(r)}dv_{CM}v^2_{CM}\times \nonumber \\
&\times \int_{v_{\text{esc}}(r)-v_{CM}}^{\sqrt{v^2_{\text{esc}}(r)-v^2_{CM}}}\frac{dv'_1}{v'_1}f_c(\sqrt{2}v_{CM})f_c(\sqrt{2}v'_1)J_{\text{sevap}}(\beta(r), \mu).
\end{align}
We show numerical calculations of $C_{\text{sevap}}$ for the Sun in Fig.~\ref{plot:SunEarthCsevap}. We note the strong exponential suppression of $C_{\text{sevap}}$ with increasing $M_{\text{DM}}$, which can be traced to the Maxwell distribution for the captured DM particles, given by Eq.~\ref{eqn:DMdistroMB}. Similar to capture and self-capture, self-evaporation features two distinct scattering regimes. The short-range regime occurs for large $M_{\gamma_D}$, when the typical momentum transfer is much smaller than the mediator mass. The interactions are effectively pointlike and result in a  $C_{\text{sevap}}\sim\mu^{-2}\sim M^{-4}_{\gamma_D}$ scaling. Long-range scattering dominates in the opposite regime, when $M_{\gamma_D}$ is small compared to the typical momentum transfer, and result in a $C_{\text{sevap}}\sim\mu^{-1/2}\sim M^{-1}_{\gamma_D}$ scaling with mediator mass.


\bibliographystyle{jhep}
\bibliography{references_list_v12}

\end{document}